\documentclass[fleqn,usenatbib]{mnras}

\usepackage[T1]{fontenc}
\usepackage{ae,aecompl}

\usepackage{graphicx}	
\usepackage{amsmath}	
\usepackage{amssymb}	
\usepackage{multicol}
\usepackage{booktabs,tabularx}
\usepackage{enumitem}
\usepackage{multirow}
\usepackage{xcolor}

\newcommand{\ee}[1]{\times 10^{#1} } 
 
\newcommand{\swift}{\textit{Swift}}
\newcommand{\hr}{\textit{HR}} 
\newcommand{\xmm}{\textit{XMM--Newton}} 
\newcommand{\chandra}{\textit{Chandra}} 
\newcommand{\nicer}{\textit{NICER}} 
\newcommand{\nustar}{\textit{NuSTAR}} 
\newcommand{\suzaku}{\textit{Suzaku}} 
\newcommand{\msun}{\text{M}_\odot}

\title[The flaring, changing-look AGN NGC~5273]{Multiple Flares in the Changing-Look AGN NGC~5273}

\author[Neustadt et al.]{
J.~M.~M.~Neustadt$^{1}$\thanks{E-mail: neustadt.7@osu.edu (JMMN)}, 
J.~T.~Hinkle$^{2}$, 
C.~S.~Kochanek$^{1,3}$,
M.~T.~Reynolds$^{4,1}$,
\newauthor{
S.~Mathur$^{1,3}$, 
M.~A.~Tucker$^{1,3}$,
R.~Pogge$^{1,3}$,
K.~Z.~Stanek$^{1,3}$,
A.~V.~Payne$^{2}$,
}
\newauthor{
B.~J.~Shappee$^{2}$,
T.~W.-S.~Holoien$^{5}$,
K.~Auchettl$^{6,7,8}$, 
C.~Ashall$^{9}$,
}
\newauthor{
T.~deJaeger$^{2}$,
D.~Desai$^{2}$, 
A.~Do$^{2}$, 
W.~B.~Hoogendam$^{2}$,
M.~E.~Huber$^{2}$
}
\\
$^{1}$Department of Astronomy, The Ohio State University, 140 West 18th Avenue, Columbus, OH 43210, USA \\
$^{2}$Institute for Astronomy, University of Hawai'i, 2680 Woodlawn Drive, Honolulu, HI 96822, USA \\
$^{3}$Center for Cosmology and AstroParticle Physics (CCAPP), The Ohio State University, 191 W. Woodruff Avenue, Columbus, OH 43210, USA \\ 
$^{4}$Department of Astronomy, University of Michigan, 1085 South University Avenue,  Ann Arbor, MI 48103, USA \\ 
$^{5}$The Observatories of the Carnegie Institution for Science, 813 Santa Barbara St., Pasadena, CA 91101, USA \\
$^{6}$Department of Astronomy and Astrophysics, University of California, Santa Cruz, CA 95064, USA \\
$^{7}$ARC Centre of Excellence for All Sky Astrophysics in 3 Dimensions (ASTRO 3D), VIC 3010, Australia \\
$^{8}$School of Physics, The University of Melbourne, Parkville, VIC 3010, Australia \\
$^{9}$Department of Physics, Virginia Tech, Blacksburg, VA 24061, USA \\
}
\date{Accepted XXX. Received YYY; in original form ZZZ}
\pubyear{2022}

\begin{document}
\label{firstpage}
\pagerange{\pageref{firstpage}--\pageref{lastpage}}
\maketitle

\begin{abstract}
NGC~5273 is a known optical and X-ray variable AGN.  We analyze new and archival IR, optical, UV, and X-ray data in order to characterize its long-term variability from 2000 to 2022.  At least one optical changing-look event occurred between 2011 and 2014, when the AGN changed from a Type~1.8/1.9 Seyfert to a Type~1.  It then faded considerably at all wavelengths, followed by a dramatic but slow increase in UV/optical brightness between 2021 and 2022.  Near-IR (NIR) spectra in 2022 show prominent broad Paschen lines that are absent in an archival spectrum from 2010, making NGC~5273 one of the few AGNs to be observed changing-look in the NIR.  We propose that NGC~5273 underwent multiple changing-look events between 2000 and 2022 -- starting as a Type~1.8/1.9, NGC~5273 changes-look to a Type~1 temporarily in 2002 and again in 2014, reverting back to a Type~1.8/1.9 by 2005 and 2017, respectively.  In 2022, it is again a Type~1 Seyfert.  We characterize the changing-look events and their connection to the dynamic accretion and radiative processes in NGC~5273, and propose that the variable luminosity (and thus, Eddington ratio) of the source is changing how the broad line region (BLR) reprocesses the continuum emission. 
\end{abstract}


\section{Introduction}\label{sec:intro}

The optical spectra of active galactic nuclei (AGNs) come in two flavors: Type~2, which show only narrow (full-width at half maximum, FWHM < 1000~km~s$^{-1}$) forbidden and permitted emission lines; and Type~1, which also have broad (FWHM $\sim$ 2000~km~s$^{-1}$) permitted emission lines, the brightest of which are the hydrogen Balmer series.  The unified model \citep{antonucci93,urry95} posits that Type~1 and Type~2 AGNs both possess broad line regions (BLRs), but the line of sight into the BLR is obscured by an outer ``dusty torus'' in Type~2s.  This is supported by the detection of infrared (IR) broad lines (e.g., \citealt{veilleux97,smith14,lafranca15,lamperti17,onori17}) or reflected and polarized optical broad lines (e.g, \citealt{miller90,kay94,heisler97,moran00,tran01}) in some, but not all, Type~2 AGNs.  Others have postulated that the BLR is non-existent at lower luminosities, implying the existence of ``True'' Type~2s \citep{panessa02,nicastro03,bian07,elitzur09,tran11,bianchi12,elitzur14,elitzur16}. 

In between Types~1 and 2, there are the ``intermediate'' Types~1.5, 1.8, and 1.9, which qualitatively describe the relative fluxes of the broad and narrow Balmer line components \citep{osterbrock81}.  In a Type~1, the broad line fluxes totally dominate over the narrow, while in a Type~1.5, the narrow fluxes are more noticeable.  In a Type~1.8, only broad H$\alpha$ and faint broad H$\beta$ are present, and in a Type~1.9, only broad H$\alpha$ is visible.  Though qualitative, Type 1.8/1.9 are distinct in that their H$\alpha$/H$\beta$ Balmer decrements are significantly higher compared to a Type~1/1.5 \citep{osterbrock81,osterbrock93}.  The physical mechanism driving this distinction has been attributed to dust in the BLR or partial obscuration of the BLR by the dusty torus (e.g., \citealt{osterbrock81}), although luminosity may also play a role (e.g., \citealt{goodrich89,trippe10,elitzur14}).  

While the unified model is effective in explaining many aspects of AGN diversity, it struggles to explain the phenomenon of changing-look AGNs, also called ``changing-state'' AGNs (see \citealt{graham20,ricci22}), where broad emission lines appear and/or disappear on timescales of years (e.g, \citealt{shappee14,lamassa15,runnoe16,macleod16,ruan16,macleod19,sheng20,ross20,jwang22}) and even sometimes months (e.g., \citealt{trakhtenbrot19,frederick19,laha22,zeltyn22}, see also ``rapid turn-on'' events, e.g., \citealt{gezari17,yan19}).  Put another way, changing-look events are when AGNs transition from one spectral type to another.  Changing-look events are coincident with significant changes in the UV/optical continuum -- as the broad lines appear/disappear, the UV/optical flux rises/fades. The X-ray fluxes can vary dramatically (e.g., \citealt{guolo21}) and in some cases be anti-correlated with the broad line and UV/optical variability \citep{trakhtenbrot19,laha22}.  Whatever the mechanism driving the changing-look event, this implies that the BLR and the UV/optical accretion disc may be affected differently than the X-ray corona.  Significant changes in absorption have been seen in the X-ray spectra of AGNs, a phenomenon which has also been called changing-look (e.g. \citealt{matt03,puccetti07,bianchi09,marchese12}), alternatively called ``changing-obscuration'' \citep{ricci22}, but these events appear to be separate from the broad emission line changing-look/changing-state phenomenon (hereafter, changing-look will refer specifically to changing-state unless stated otherwise).  For most changing-looks, variable obscuration along the line of sight is a poor explanation \citep{macleod16}, though there are exceptions \citep{zeltyn22}. Rather, the overall luminosity and mass-inflow rate of these AGNs seem to be changing on much shorter timescales than those predicted by conventional accretion disc theory \citep{shakura73}.  

Transient phenomena further complicate AGN studies.  Tidal disruption events (TDEs, see \citealt{gezari21} for an overview) change the mass-inflow rate through the rapid accretion of disrupted stellar material, and some changing-look events have been proposed to be TDEs-in-AGNs \citep{merloni15,blanchard17,ricci20,ricci21}. Nevertheless, these events are short-lived in the UV/optical compared to changing-looks \citep{macleod19}.  There is also the ``class'' of ambiguous nuclear transients (ANTs) which fall into the overlap of rapid changing-looks, TDEs, and supernovae occurring in galactic nuclei (e.g., \citealt{neustadt20,holoien22,yu22,hinkle22-18el,hinkle22-20hx,hinkle22-ants}).  How these events fit into AGN variability is unclear, but the overlap of all of these phenomena hint at an underlying ``continuum'' of events defined by the changes in accretion onto the supermassive black hole (SMBH) and how those changes propagate outward into the surrounding galactic nucleus.

Here we discuss the Seyfert galaxy \citep{seyfert43} NGC~5273.  Located at redshift $z=0.00362$, the distance to the galaxy has been measured to be $D = 16.5 \pm 1.6 \rm ~Mpc$ using surface brightness fluctuations (\citealt{tonry01}, updated by \citealt{tully08}). The SMBH mass was measured using reverberation mapping (RM, \citealt{bentz14}) to be $M_{\rm BH} = (4.7 \pm 1.6) \ee{6} ~\msun$. This is one of the least massive AGNs to have an RM mass estimate.  It is also relatively dim compared to other AGNs, with $L_{\rm X}/L_{\rm Edd} \sim$~10$^{-4}$--10$^{-3}$ \citep{papadakis08,vincentelli20}, where $L_{\rm X}$ is the 2--10~keV X-ray luminosity, and $L_{\rm Edd}$ is the Eddington luminosity.  As such, it has been often classified as a low-luminosity AGN (LLAGN).

There are at least five studies of the broad emission lines of NGC~5273 between 1980 and 2010. \citet{stauffer82} noted that the broad H$\alpha$ flux varied by a factor of 2.3 in their observations between 1980 and 1981 with the Lick 3-m telescope.  \citet{dahari88} classified NGC~5273 as a Type~1 Seyfert based on the \citet{stauffer82} spectra, but this classification is puzzling because no broad H$\beta$ flux is reported (see Tab.~1 in \citealt{dahari88}). \citet{ho95,ho97} classified it as a Type~1.5 based on the faint but visible broad H$\beta$ and H$\gamma$ emission lines in their 1984 Palomar spectrum.  By contrast, \citet{osterbrock93} classified it as Type~1.9 using a spectrum from 1991 taken with the Lick 3-m, measuring a H$\alpha$/H$\beta$ Balmer decrement of 10.9 (6.3 after correcting for reddening). \citet{trippe10} re-classified it as a Type~1.5 based on a 2008 KPNO spectrum, though they argued that the difference between the \citet{osterbrock93} classification and theirs was due to host galaxy subtraction and not due to intrinsic changes in the spectra.  This assertion is also puzzling, as both analyses included host subtraction, and, by eye, the H$\beta$ profiles in the two spectra look quite different (Fig.~3 in \citealt{osterbrock93}, Fig.~A1, Panel 10 in \citealt{trippe10}).  Because of the broad range of instruments and lack of access to the spectra (other than the \citealt{ho95,ho97} spectrum), it is difficult to quantitatively compare these results, but they are suggestive of previous short-term structural changes in the spectra of NGC~5273.

When NGC~5273 was observed as part of the Sloan Digital Sky Survey (SDSS, \citealt{ahumada20}) in 2006, it was clearly a Type~1.8/1.9 due to its weak broad H$\beta$ flux.  When next observed in 2014 for the \citet{bentz14} RM campaign, NGC~5273 was a quintessential Type~1 Seyfert with strong broad H$\beta$, H$\gamma$, and other Balmer features, indicating that NGC~5273 underwent a changing-look event sometime between 2006 and 2014.  While the H$\beta$ line was used for the RM campaign, \citet{bentz14} did not remark on the changing-look.  We could find no optical spectra of NGC~5273 taken between 2014 and 2022 in published articles or public archives.

NGC~5273 began to steadily increase in brightness in late 2021, peaking in 2022.  This was observed by the All-Sky Automated Survey for Supernove (ASAS-SN, \citealt{shappee14,kochanek17}), and a follow-up spectrum was taken by the Spectroscopic Classification of Astronomical Transients survey (SCAT, \citealt{tucker22}).  The new spectrum showed significant changes in the broad Balmer lines from the archival SDSS spectrum, leading ASAS-SN and SCAT reporting NGC~5273 as a transient \citep{atel}.  As the past and present evolution of NGC~5273 became clearer, we initiated a follow-up campaign using various ground- and space-based instruments.  Here we report our results showing NGC~5273 to be a changing-look AGN, with strong evidence for multiple changing-look events occurring over the past two decades.  In Section~\ref{sec:observations}, we discuss the details of the follow-up campaign and the archival data we use to chart the evolution of NGC~5273.  In Sections~\ref{sec:phot} and \ref{sec:spec}, we discuss the photometric and spectroscopic evolution, respectively, of the AGN.  In Section~\ref{sec:xray}, we analyze the evolution of the X-ray properties, and we discuss our findings in Section~\ref{sec:discuss}.  We briefly summarize the results in Section~\ref{sec:conc}.

\begin{figure*}
\includegraphics[width=\linewidth]{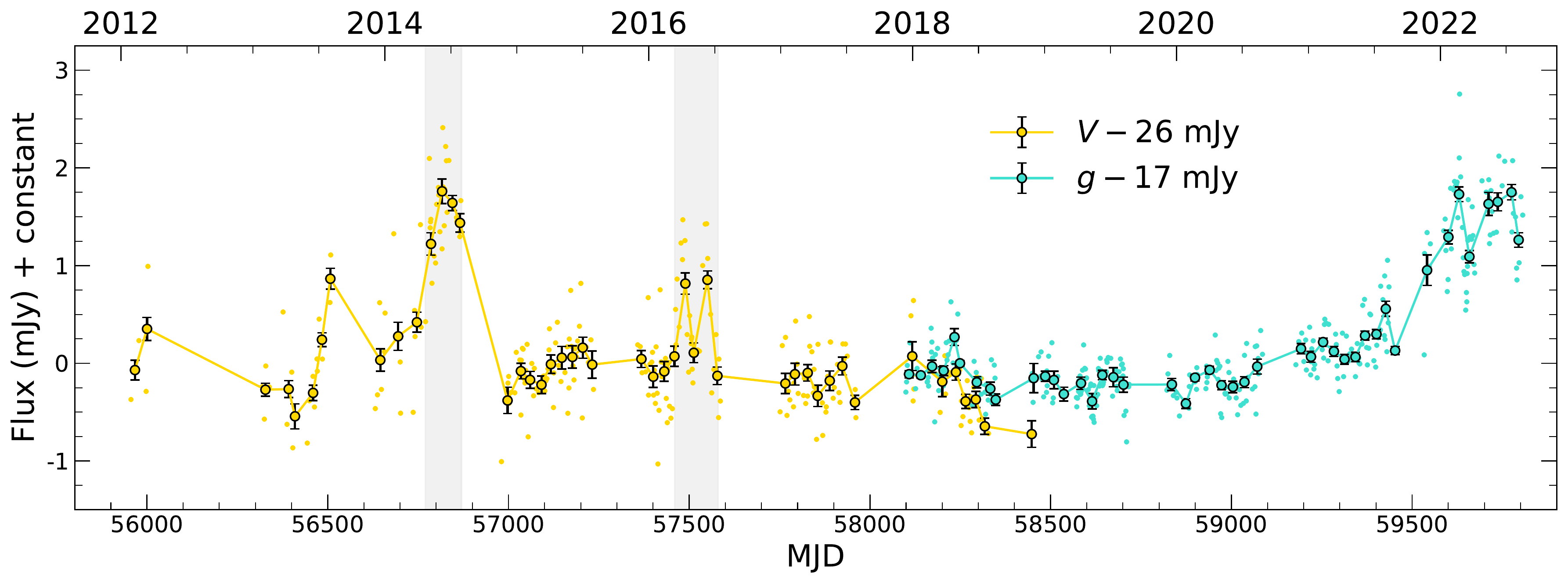}
\caption{ASAS-SN $V$-band (yellow) and $g$-band (turquoise) lightcurves of NGC~5273.  Dots are the individual epochs, and the filled circles with errors are a rolling 28~d mean flux.  The median flux (26~mJy for $V$, 17~mJy for $g$) has been subtracted to highlight the flares in the lightcurves.  In addition to the slow rise beginning in $\sim$2021, we also highlight two flares in the lightcurve in mid-2014 and mid-2016.}
\label{fig:asassn-lc}
\end{figure*}


\section{Observations}\label{sec:observations}

In this section, we summarize the available archival data and our new photometry and spectroscopy of NGC~5273.  As discussed in Section~\ref{sec:intro}, the photometric and spectroscopic variability of NGC~5273 has been tracked for decades.  For the purposes of our analysis, we focus on the data only as far back as 2000 (with the exception of the 1984 Palomar spectrum) where we begin to have overlapping X-ray, UV, optical, and IR photometric and spectroscopic data. 

\subsection{ASAS-SN photometry}

\begin{figure*}
\includegraphics[width=\linewidth]{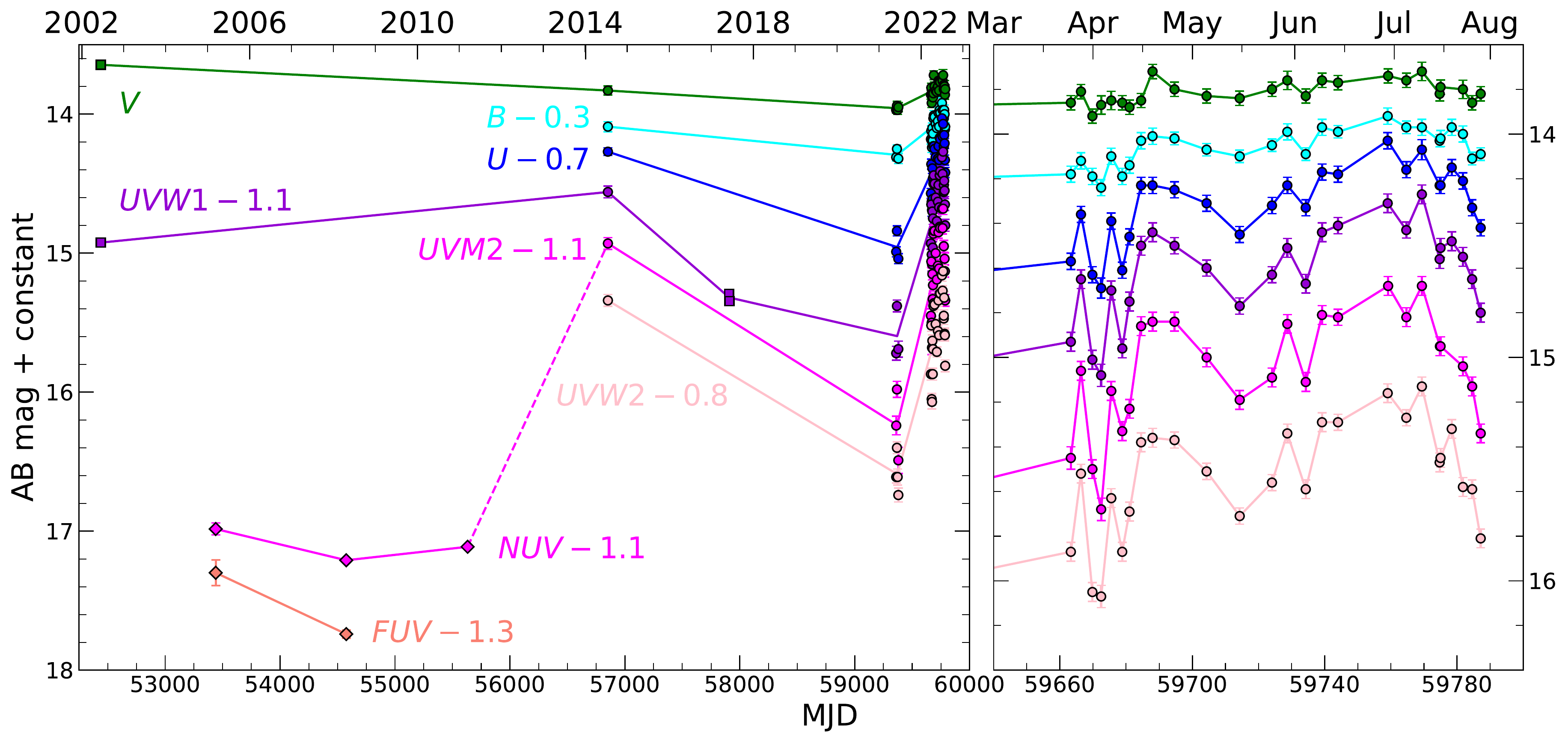}
\caption{ \textbf{Left:} Lightcurves of NGC~5273 from 2002 to 2022 in the \swift~UVOT (circles) and \textit{XMM}-OM UV/optical (squares) filters  and GALEX UV filters (rhombuses).  While not identical, the GALEX \textit{NUV} filter is very similar to the \textit{UVM}2 filter, indicating a significant change in flux between 2005--11 and 2014. \textbf{Right:}  \swift~UVOT lightcurves from 2022.  Note that the vertical scale of the right panel is significantly compressed relative to the left panel.}
\label{fig:swift-lc}
\end{figure*}

\begin{figure}
\includegraphics[width=0.95\linewidth]{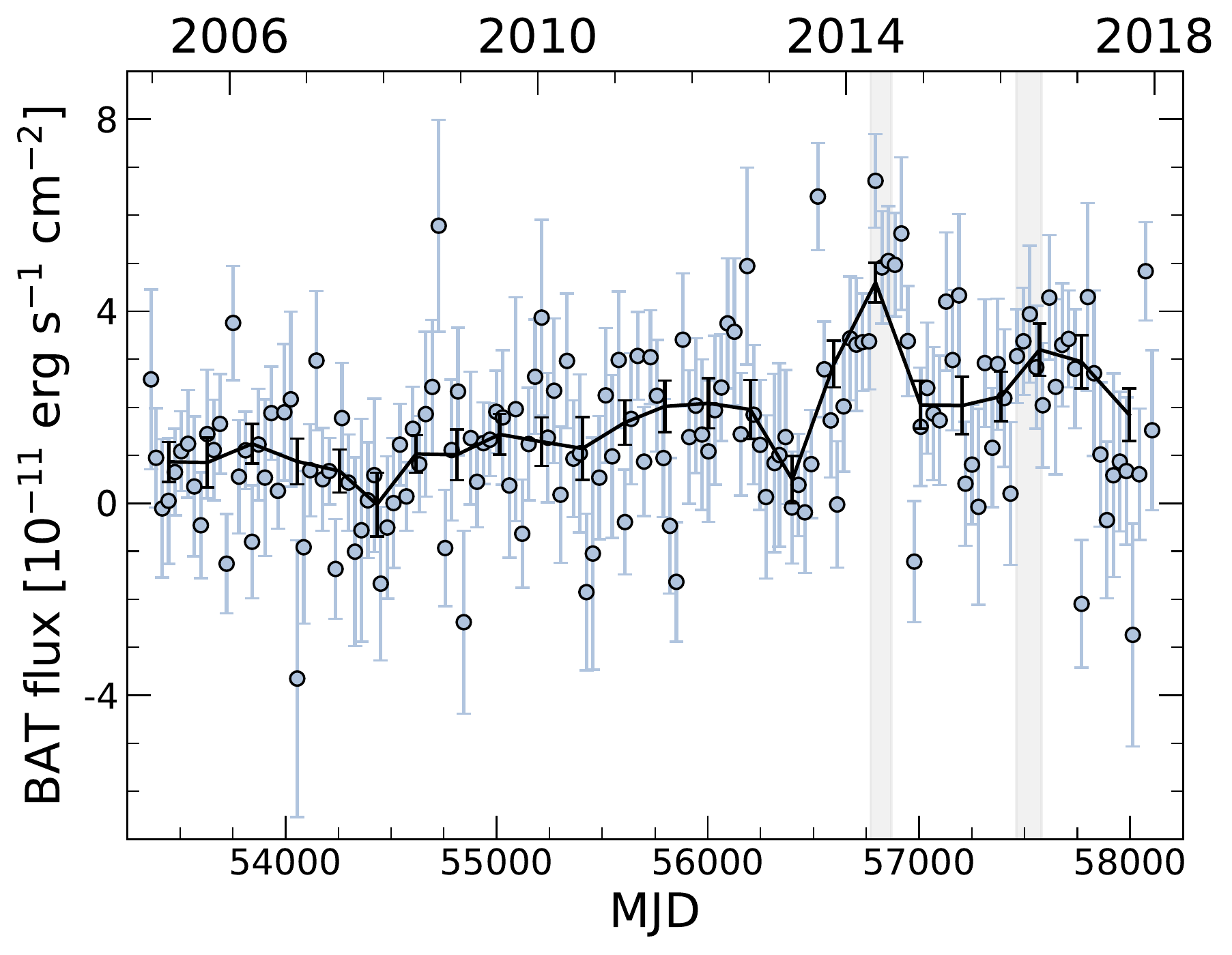}
\caption{\swift~BAT 14--195~keV X-ray lightcurve of NGC~5273.  Monthly binned data are shown as purple points, and the 6-month binned error-weighted means are shown as a black line.  The grey shaded regions correspond to the 2014 and 2016 flares seen in the ASAS-SN lightcurve (see Fig.~\ref{fig:asassn-lc}).}
\label{fig:bat}
\end{figure}

\begin{figure*}
\includegraphics[width=\linewidth]{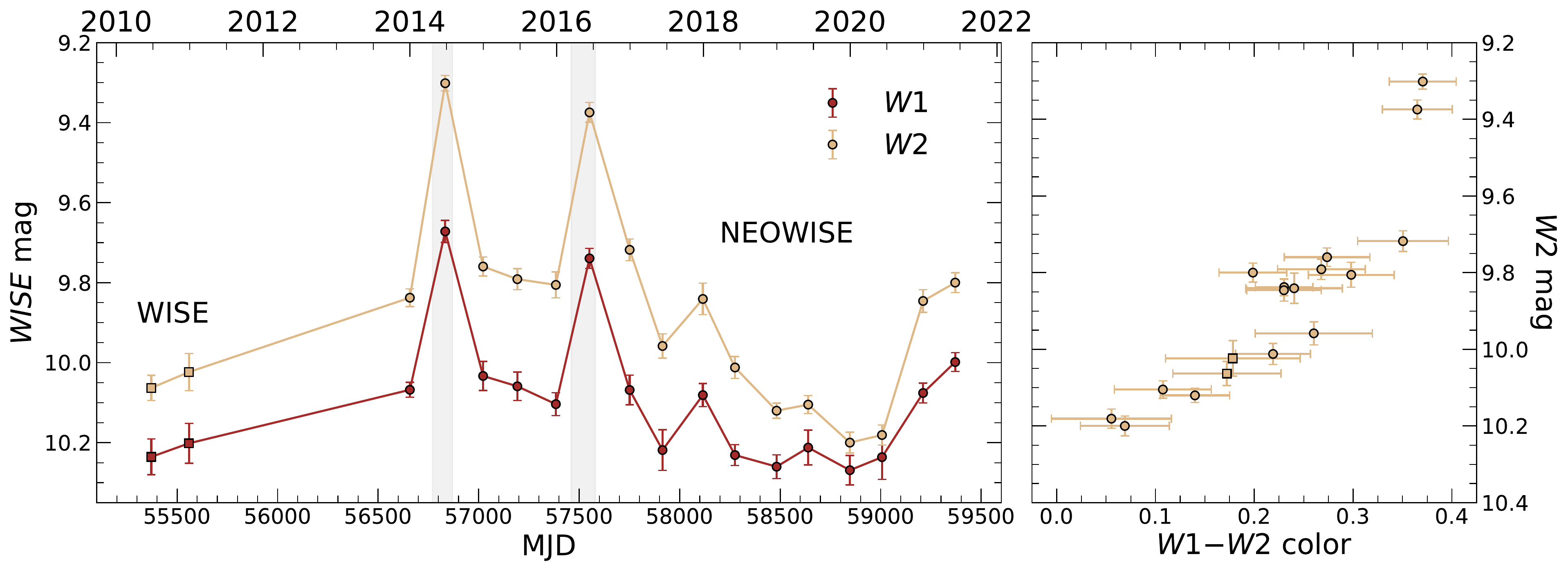}
\caption{\textbf{Left:} WISE (squares) and NEOWISE (circles) lightcurves of NGC~5273. The grey shaded regions correspond to the 2014 and 2016 flares seen in the ASAS-SN and \swift~BAT lightcurves (see Figs.~\ref{fig:asassn-lc} and \ref{fig:bat}). \textbf{Right:} $W1$--$W2$ as a function of the $W2$ magnitude.   The color uncertainty is defined as the quadratic sum of the $W1$ and $W2$ mean absolute errors.  A positive correlation between color and flux is clearly present.}
\label{fig:wise}
\end{figure*}

The ASAS-SN data reduction and image subtraction procedures are described in \citet{shappee14} and \citet{kochanek17}.  We subtract out the median difference between the fluxes with the same filter from different ASAS-SN detectors to account for systematic offsets between detectors.  The $V$- and $g$-band ASAS-SN lightcurves are presented in Figure~\ref{fig:asassn-lc}.  

\subsection{\swift\ data}

We obtained follow-up observations with the \textit{Neil Gehrels Swift Observatory} (\swift, Target ID: 15085, PI: J.~Neustadt) including the full suite of filters on the UltraViolet and Optical Telescope (UVOT, \citealt{roming05}) -- $V$ (5468~\AA), $B$ (4392~\AA), $U$ (3465~\AA), \textit{UVW}1 (2600~\AA), \textit{UVM}2 (2246~\AA), and \textit{UVW}2 (1928~\AA) -- along with simultaneous X-Ray Telescope (XRT, \citealt{burrows05}) observations in Photon-Counting (PC) mode.  The observations have a wide range of total exposure times, but generally, the later observations have longer exposure times to better constrain the X-ray flux.  We also analyzed other \swift\ exposures obtained from 2014 to 2021 (Target ID: 80685, PI: F.~Vincentelli), as well as the observations in 2022 carried out as part of a separate campaign (Target ID: 15085, PI: M.~Guolo).  The full \swift\ observations are listed in Table~\ref{tab:swift} and the full \swift\ UVOT lightcurves are shown in Figure~\ref{fig:swift-lc}. 

Each epoch of UVOT data consists of 2 observations in each filter. We combined the exposures and extracted counts from a 5$''$ radius around the source and from a background region with no sources and a radius of 100$''$.  For these two steps, we used the \textsc{HEASoft} v6.30 software tasks \textsc{uvotimsum} and \textsc{uvotsource}, respectively. The counts were then converted to AB magnitudes and fluxes using the UVOT \textsc{CALDB} v20211108 \citep{poole08,breeveld10}.

All XRT observations were reduced following the standard \swift\ XRT data reduction guide\footnote{http://swift.gsfc.nasa.gov/analysis/xrt\_swguide\_v1\_2.pdf}, and reprocessed using the \textsc{HEASoft} v6.30 \swift\ \textsc{xrtpipeline} v0.13.2. Standard filters and screening were applied, along with the calibration files from the XRT \textsc{CALDB} v20211108.  We used a source region centered on the position of NGC~5273 with a radius of $50''$ and a source free background region centered at ($\alpha,\delta$) $=$ (13:41:32.65,+35:38:34.61) with a radius of $175''$.  We extracted a source and background spectrum using \textsc{xrtproducts}. Ancillary response files for each spectra were generated using the task \textsc{xrtmkarf}, along with the standard response matrix files from the XRT \textsc{CALDB}.  Due to the low number of counts in many of our spectra, we binned the spectra to have a minimum of 5 counts per energy bin.

\subsection{\nicer\ data}
NGC~5273 was observed by the \textit{Neutron Star Interior Composition ExploreR} (\nicer, \citealt{nicer16}) on six occasions between 2022 April 1 and 2022 July 4 (MJD~59670--59764). The archived \nicer\ data was obtained from the \textsc{HEASARC} and processed using the \textsc{HEASoft} v6.30.1 software release and the calibration files from the \nicer\ \textsc{CALDB} v20210707. The event files were reprocessed with \textsc{nicerl2} and spectra were extracted from the cleaned event files with \textsc{xselect}. Response files were generated for each observation with the \textsc{nicerarf} and \textsc{nicerrmf} tasks. The `3C50' method was adopted to estimate the background \citep{rem22}.  We binned the spectra to have a minimum of 20 counts per energy bin.

\subsection{\xmm\ data}
We downloaded the archival \textit{X-Ray Multi-Mirror Mission} (\xmm, \citealt{jansen01}) observations from  2002 June 14 (MJD~52440), 2017 June 2 (MJD~57907), and 2017 June 5 (MJD~57909) from the \xmm\ Science Archive\footnote{http://nxsa.esac.esa.int/nxsa-web/}.  The data were reduced as part of 4\textit{XMM}-DR12 \citep{webb20}.  For simplicity, we use only the data from the EPIC-pn detector \citep{struder01} with the most recently updated response files\footnote{https://www.cosmos.esa.int/web/xmm-newton/epic-response-files}.  The spectra were binned to have a minimum of 20 counts per energy bin.

NGC~5273 was simultaneously observed with the Optical Monitor (\textit{XMM}-OM, \citealt{mason01}) instrument onboard \xmm.  The UV/optical photometry in various filters was extracted as part of the \textit{XMM}-SUSS5.0 \citep{page12,page21}.  The filters on \textit{XMM}-OM are identical to those on \swift\ UVOT, and thus we include the \textit{XMM}-OM data alongside the \swift\ data in Figure~\ref{fig:swift-lc} and Table~\ref{tab:swift}.

\subsection{\chandra\ data}
NGC~5273 was observed by the \chandra\ \textit{X-ray Observatory} (\chandra) on  2000 September 3 (MJD~51791, Obs ID: 415, PI: Garmire)\footnote{https://doi.org/10.25574/00415}. We downloaded the most recently re-processed data  (2021/03/16) from the \textsc{HEASARC}.  We followed the standard reduction pipeline for a point-like source as described in the relevant \chandra\ Analysis Guide\footnote{https://cxc.cfa.harvard.edu/ciao/threads/pointlike/} using \textsc{ciao} v4.14 \citep{fruscione06} and \textsc{caldb} v20220324.  We used a source extraction region of 10'' and a background annulus extending from 20'' to 50'' around the source.  Because of the low counts, we binned the spectra to have a minimum of 5 counts per energy bin.  

\subsection{\suzaku\ data}
\suzaku\ \citep{mitsuda07} observed NGC~5273 on 2013 July 16 (MJD~56489).  We analyze the 0.3--10~keV data obtained with the X-ray Imaging Spectrometers (XIS, \citealt{koyama07}).  We followed the standard reduction pipeline as described in the ABC guide\footnote{https://heasarc.gsfc.nasa.gov/docs/suzaku/analysis/abc/} using \textsc{HEAsoft} v6.30.1 and the \suzaku\ \textsc{CALDB} v20181010.  We generated the response matrix and ancillary response files with \textsc{xisrmfgen} and \textsc{xissimarfgen} \citep{ishisaki07}. We used a 260'' source extraction region, and we combined the spectra of the individual XIS detectors.  We binned the spectra to have a minimum of 100 counts per energy bin.

\subsection{X-rays above 10 keV}

There are observations from 2003 January to 2021 August with the \textit{INTErnational Gamma-Ray Astrophysics Laboratory} (\textit{INTEGRAL}) at 20--100~keV.  The \textit{INTEGRAL} data from 2003 to 2014 July were analyzed in \citet{pahari17}, and while they observed variability, there were no coherent changes over this period.   In 2014, roughly coincident with the \citet{bentz14} RM campaign, NGC~5273 was observed with the \textit{Nuclear Spectroscopic Telescope Array} (\textit{NuSTAR}), and the resulting spectrum was analyzed in \citet{pahari17} to find a power-law index of $\Gamma \sim 1.8$ and a cut-off energy of $\sim$140~keV, which are typical of AGNs \citep{ricci17}. 

NGC~5273 was observed from 2003 December to 2017 December as part of the 157-month \textit{Swift}-BAT survey\footnote{https://swift.gsfc.nasa.gov/results/bs157mon/} (Lien et al. in prep.).  We downloaded the Crab-weighted monthly-binned \swift~BAT lightcurve and converted this to physical units, as shown in Figure~\ref{fig:bat}. 

\subsection{GALEX data}

The Galaxy Evolution Explorer (GALEX, \citealt{martin2005, morrissey2005, morrissey2007}) operated from 2000--2013 and simultaneously observed in two UV filters: \textit{NUV} (2305~\AA) and \textit{FUV} (1549~\AA). 3 epochs of archival GALEX \textit{NUV} photometry (2 for \textit{FUV}) were obtained using the \textsc{gPhoton} package \citep{million2016}. Fluxes were measured within a 5\arcsec\ aperture and the background flux was estimated using an annulus extending from 60--120\arcsec. The background annulus may contain some of the host-galaxy light, so we checked that the host does not contaminate the derived photometry by comparing our aperture photometry to the GALEX Merged Catalog of Sources (MCAT) photometry, which estimates the background flux using $192\arcsec\times192\arcsec$ spatial bins \citep{morrissey2007}. The two sources of photometry agree to $<1\sigma$ with a maximum difference of $0.02\pm0.05$~mag.  The GALEX photometry is shown in Figure~\ref{fig:swift-lc} and Table~\ref{tab:swift}.

\subsection{Mid-infrared photometry}

NGC~5273 was observed by the Wide-field Infrared Survey Explorer (WISE, \citealt{wise}) and by the Near-Earth Object Wide-field Infrared Survey Explorer (NEOWISE, \citealt{neowise1,neowise2}) in the $W1$ ($3.4~\rm \mu m$) and $W2$ ($4.6~\rm \mu m$) filters.  The mid-IR (MIR) data were downloaded from NASA/IPAC Infrared Science Archive\footnote{https://irsa.ipac.caltech.edu/frontpage/}.  There are 2 epochs of WISE data from 2010, and there are 16 of NEOWISE spanning from late 2013 to 2021. For each epoch, there are multiple exposures separated by small time intervals ranging from $\sim$1~hr to $\sim$1~d, with slightly varying flux measurements.  We calculated the mean flux for each epoch, with the errors calculated as the mean absolute error. These are shown in Figure~\ref{fig:wise}. 

\begin{figure*}
\includegraphics[width=\linewidth]{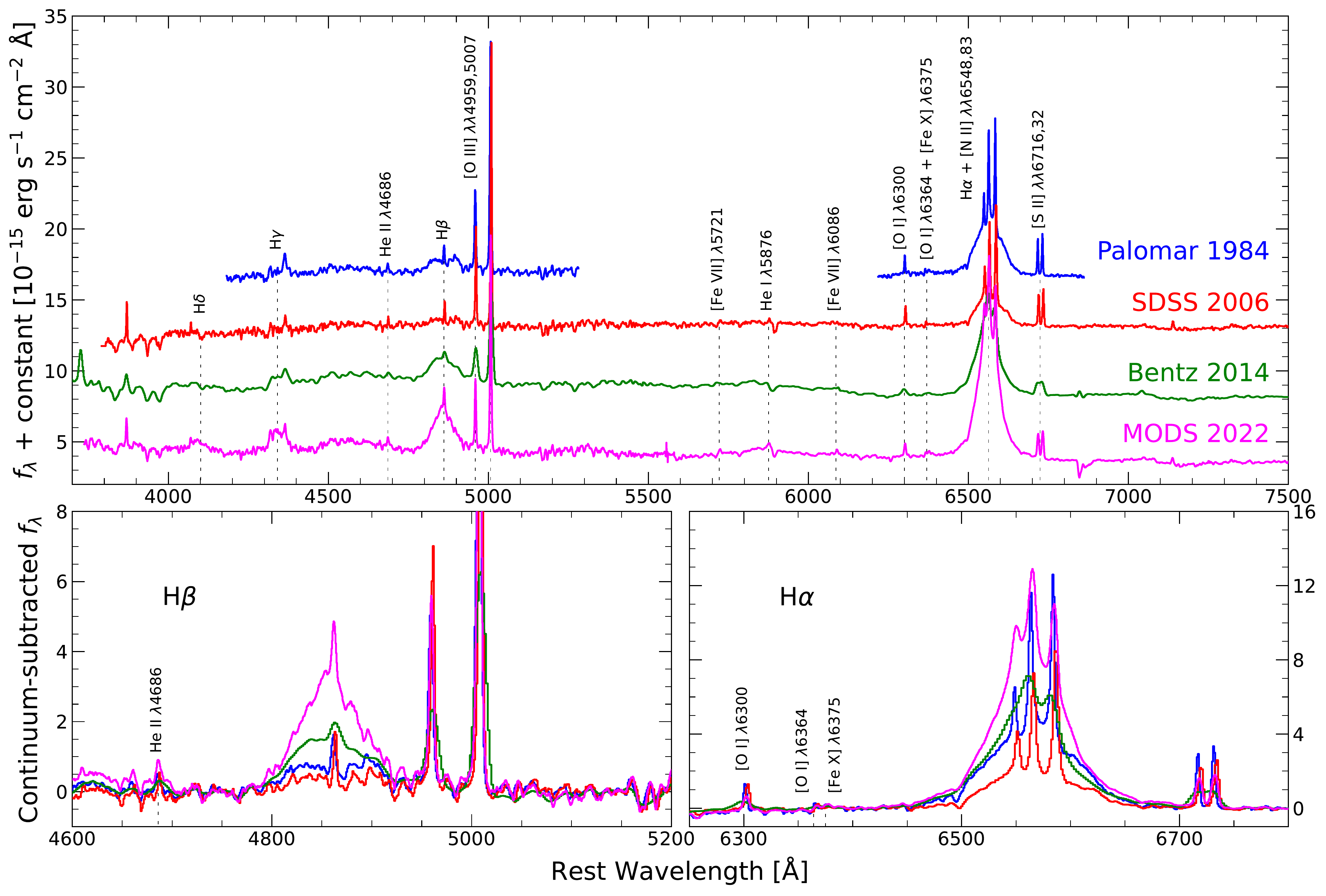}
\caption{Spectroscopic evolution of NGC~5273 since 1984.  Spectra are normalized such that the integrated fluxes of the [\ion{S}{ii}] and [\ion{O}{iii}] narrow emission lines are constant and fixed to the SDSS values.  Prominent spectral features are labelled.  \textbf{Top:} The full spectra.  \textbf{Bottom Left:} The H$\beta$ profiles.  \textbf{Bottom Right:} The H$\alpha$ profiles.  The broad Balmer fluxes are much stronger in 2014 and 2022 than in 1984 or 2006.  Especially noteworthy is the change in the broad H$\beta$ flux, going from negligible in 1984 and 2006 to ``booming'' in 2014 and 2022.  The changes in H$\beta$ are the changing-look from Type~1.8/1.9 to Type~1.}
\label{fig:spec_archive}
\end{figure*}

\subsection{Optical spectroscopy}

We obtained multi-epoch optical spectra of NGC~5273 spanning 164~d from 2022 March 11 until 2022 August 22 (MJD~59649--59813) using the Low-Resolution Imaging Spectrometer (LRIS, \citealt{oke95}) on the 10-m Keck I telescope, the Multi-Object Double Spectrographs (MODS, \citealt{pogge10}) on the dual 8.4-m Large Binocular Telescope (LBT), and the Supernova Integral Field Spectrograph (SNIFS, \citealt{lantz04}) on the University of Hawaii 88-in telescope (UH88). The SNIFS data reduction and calibration procedures are described in \citet{tucker22}.  The Keck and MODS data were reduced using standard procedures, with bias and flat field corrections, wavelength solutions from arc lamps, and spectrophotometric standard stars from the same nights as the observations.  The Keck reductions were done with aid of \textsc{PypeIt} \citep{pypeit} and the MODS reduction with \textsc{modsIDL} \citep{modsidl}.

We also obtained archival optical spectra from the \citep{ho95,ho97} observations taken with the Palomar Hale 200" telescope \citep{oke82,ho95}, SDSS Data Release 16 (DR16, \citealt{ahumada20}), and the \citet{bentz14} RM campaign (hereafter referred to as the ``Bentz'' spectrum) that used the 3.5-m telescope at Apache Point Observatory (APO).  We show these archival spectra along with one of the 2022 MODS spectra in Figure~\ref{fig:spec_archive}.  The new spectra are shown in Figure~\ref{fig:spec_full}.  The relevant details, such as slit width and spectral resolution, of both the new and archival spectroscopic observations are listed in Table~\ref{tab:spec}.

\subsection{Near-infrared spectroscopy}

We obtained three new near-IR (NIR) spectra of NGC~5273 using SpeX \citep{rayner03} on the NASA Infrared Telescope Facility (IRTF). The spectra were reduced with \textsc{Spextool} \citep{cushing04} using flat and arc lamps taken after the science spectra and flux calibrated using nearby A0 telluric standard stars within a typical airmass difference of 0.05. We also reduced an archival IRTF spectrum from 2010 \citep{lamperti17} using the same methods as the new spectra.  These spectra are shown in Figure~\ref{fig:irspec} and included in Table~\ref{tab:spec}.


\section{Photometric evolution}\label{sec:phot}

In Figure~\ref{fig:asassn-lc}, there are two noticeable flares in the $V$-band in 2014 and 2016, with the 2014 flare being coincident with the \citet{bentz14} RM campaign.  Starting in 2021, NGC~5273 begins to slowly increase in brightness in the $g$-band, reaching a maximum around 2022 February 21 (MJD~59631).  This behavior is very different from the fast rises ($\lesssim$ 100~d) of the 2014 and 2016 flares, though this could be due to the relative lack of data during the earlier flares.  While the ASAS-SN lightcurve shows these distinct features, these flares and bright periods still represent relatively small changes in the total $V$- and $g$-band flux, of order 10~per~cent.  This is due to the overall optical dimness of the AGN relative to the host galaxy, as NGC~5273 is often characterized as a LLAGN.  \citet{trippe10} estimated that 80~per~cent of the flux at 5100~\AA\ is from the host galaxy, and a similar value is calculated by \citet{bentz14}.  The overall changes in flux are much more prominent in the UV. 

As shown in Figure~\ref{fig:swift-lc}, we do not have the same UV coverage for each epoch.  Between 2005--2011, we have GALEX observations rather than \swift~UVOT or \textit{XMM}-OM, and in 2002 and 2017, we have only \textit{UVW}1 observations rather than the full \textit{XMM}-OM filter set.  However, based on the similar effective central wavelengths of the GALEX \textit{NUV} and \textit{UVM}2 filters ($\sim$2300~\AA), we can interpret these as being roughly equivalent.  We can also estimate the \textit{UVM}2 flux at the time of the 2002 and 2017 \textit{UVW}1 observations by assuming a \textit{UVW}1--\textit{UVM}2 color of $\sim$0.4~mag based on the 2014 and 2021--2022 observations.  From this, we can say that the predicted \textit{UVM}2 flux in 2002 would be at least 1~mag brighter than observed in 2005--2011.  Furthermore, there is a 2~mag difference between the predicted 2005--2011 \textit{UVM}2 flux and the observed 2014 flux -- this UV-bright flare is coincident with the \citet{bentz14} RM campaign and the optical flare seen in the ASAS-SN lightcurve in Figure~\ref{fig:asassn-lc}.  After 2014, the flux drops.  It is 0.8--1.2~mag fainter in \textit{UVW}1 in 2017--2021 and 1.0--1.5~mag fainter in \textit{UVM}2 in 2021 than in 2014. It is, however, brighter in the UV in 2017--2021 than in the 2005--2011 observations by 0.5--1.0~mag.  In other words, the low-flux state between 2014 and 2022 is brighter than the low flux state between 2002 and 2014.

In the \swift~BAT lightcurve shown in Figure~\ref{fig:bat}, it is clear that the 2014 and 2016 flares seen in the ASAS-SN lightcurve in Figure~\ref{fig:asassn-lc} are present in the X-rays as well.  The duration of these flares appears longer in the X-rays than in the optical, though this may be due to the lack of full temporal coverage in the ASAS-SN lightcurve due to seasonal gaps.  We also observe a slow increase in the 6-month binned BAT flux from late-2004 to 2017, roughly doubling the overall flux.  This is likely related to the increase in UV flux from 2005--2011 until 2017--2021 seen in Figure~\ref{fig:swift-lc} and discussed above.  Besides this slow increase, it appears quiescent from 2005--2011, matching the UV data.  

Looking at the WISE and NEOWISE lightcurves in Figure~\ref{fig:wise}, we again see the 2014 and 2016 flares that occur in the ASAS-SN and \swift~BAT lightcurves in Figures~\ref{fig:asassn-lc} and \ref{fig:bat}, respectively, also occur in the MIR.  We also observe that as the MIR flux increases, the $W1$--$W2$ color becomes redder.  It is always bluer than the $W1$--$W2>0.8$ color of typical luminous AGNs \citep{stern12,assef13}, presumably because there is still a significant flux contribution from the host.  Thus, as the AGN gets brighter relative to the constant host flux, the color becomes more ``AGN-like''.


\begin{figure*}
\includegraphics[width=\linewidth]{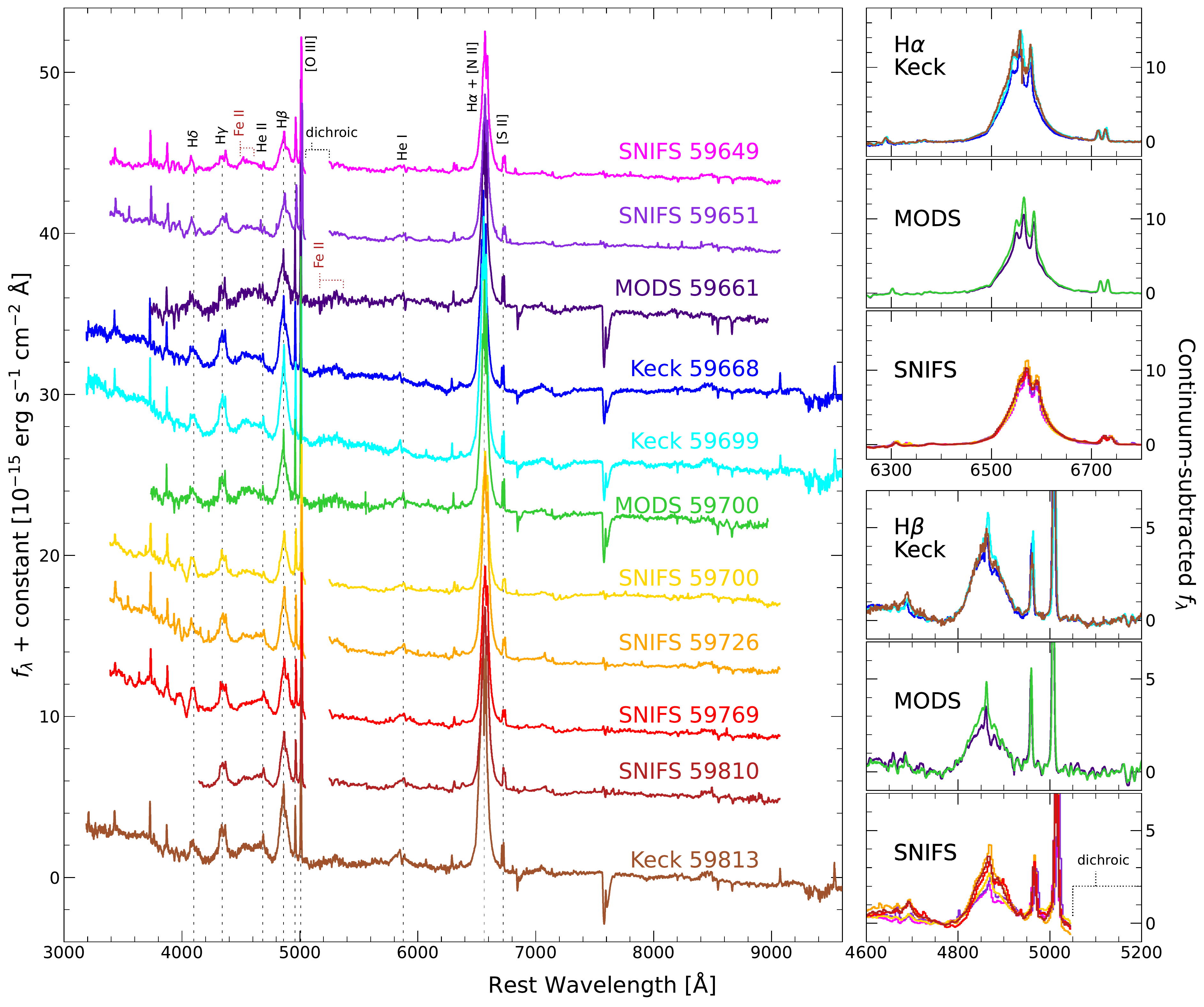}
\caption{Spectroscopic evolution of NGC~5273 over our follow-up observing campaign.  The observations are ordered by date, increasing from top to bottom (violet to brown).  Spectra are normalized such that the integrated fluxes of the [\ion{S}{ii}] and [\ion{O}{iii}] narrow emission lines are constant and fixed to the SDSS values.  Prominent spectral features are labelled.  The dichroic overlap regions of the SNIFS spectra have been excised. The SNIFS spectra are corrected for telluric absorption, the Keck and MODS are not. \textbf{Left:} The full spectra.  \textbf{Top Right:} The H$\alpha$ profiles.  \textbf{Top Right:} The H$\beta$ profiles.  The Balmer profiles have been separated by instrument so as to accurately compare changes over time.}
\label{fig:spec_full}
\end{figure*}

\begin{figure*}
\includegraphics[width=\linewidth]{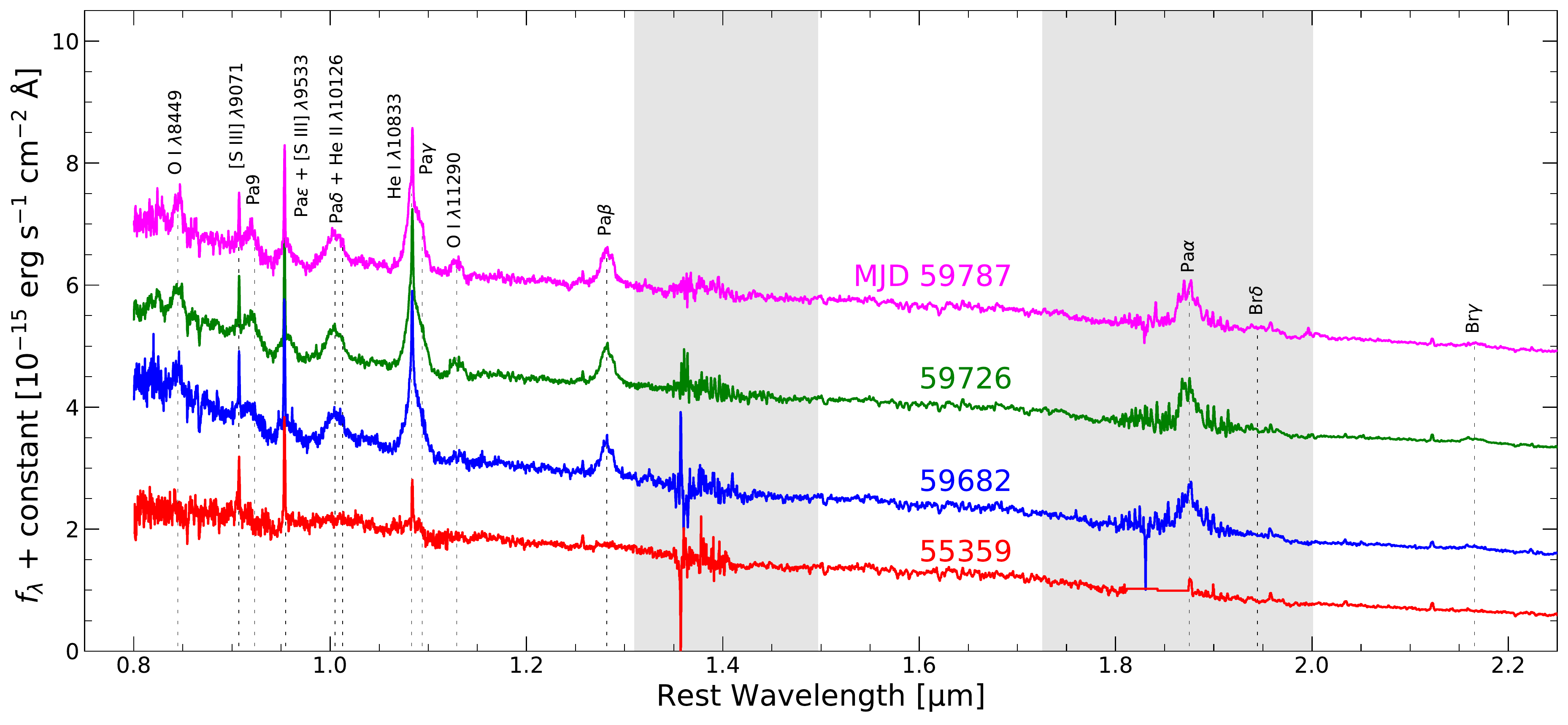}
\caption{NIR spectra of NGC~5273 showing the appearance of broad lines coincident with the 2022 optical changing-look event.  The grey bands correspond to telluric features where the spectra are less reliable.  The lack of broad lines in the archival 2010 spectrum points towards NGC~5273 being a ``True'' Type~1.8/1.9 rather than being obscured/dust-reddened (see Sec.~\ref{sec:spec}).  }
\label{fig:irspec}
\end{figure*}

\section{Spectroscopic evolution}\label{sec:spec}

Between 2006 and 2014, NGC~5273 undergoes a changing-look event from a Type~1.8/1.9 to Type~1 as evidenced by the differences in the spectra shown in Figure~\ref{fig:spec_archive}.  Where the 2006 SDSS spectrum has a very faint broad H$\beta$ component with flux comparable to the narrow H$\beta$ component, the 2014 Bentz spectrum show strong broad H$\beta$ and other Balmer lines with fluxes that dwarf their associated narrow lines.  As mentioned in Section~\ref{sec:intro}, \citet{bentz14} used the broad H$\beta$ line for their RM campaign but did not remark on the changes from the earlier spectra.  As discussed in Section~\ref{sec:intro}, NGC~5273 has been variously typed as a Type~1.8/1.9 and Type~1/1.5 Seyfert over the decades \citep{dahari88,osterbrock93,ho95,trippe10}.  While most of these classifications were qualitative, it is clear that the 1984 Palomar and 2006 SDSS spectra both have significantly weaker broad H$\beta$ and H$\gamma$ than the 2014 spectrum.  

In addition to the broad and narrow Balmer lines, we identify common narrow lines observed in AGN spectra, including: \ion{He}{ii}~$\lambda$4686, [\ion{O}{iii}]~$\lambda\lambda$4959,5007, [\ion{Fe}{vi}]~$\lambda$5571, \ion{He}{i}~$\lambda$5876, [\ion{Fe}{vi}]~$\lambda$6086, [\ion{O}{i}]~$\lambda\lambda$6300,64, [\ion{Fe}{x}]~$\lambda$6375, [\ion{N}{ii}]~$\lambda\lambda$6548,83, and [\ion{S}{ii}]~$\lambda\lambda$6716,32.

We simultaneously fit the H$\alpha$ region with narrow H$\alpha$, [\ion{N}{ii}]~$\lambda\lambda$6548,83, and [\ion{S}{ii}]~$\lambda\lambda$6716,32 lines on top of a broad Gaussian H$\alpha$ component.  The best-fit values and errors are computed using \textsc{scipy.optimize.curve$\_$fit} \citep{scipy}.  Because the spectra were taken with different instruments and different resolutions, we scale the fluxes to keep the integrated fluxes of the [\ion{S}{ii}] lines constant and set to the fluxes measured in the SDSS spectrum.  By comparing the two lines of the [\ion{S}{ii}] doublet, we estimate that the scalings between spectra are consistent to $\sim$10~per~cent. Even with the very different slit widths (1--5'', see Tab.~\ref{tab:spec}), the scaling factors were often of order unity, meaning that the significant changes we see in the broad flux are almost certainly real rather than artificial.  This also implies that there are no significant changes in the narrow line fluxes over time.  For the SDSS spectrum, we calculate the error on the H$\alpha$ flux from the estimated errors of the broad Gaussian fit parameters.  For the H$\beta$ region, we repeat the process using narrow H$\beta$ and [\ion{O}{iii}]~$\lambda\lambda$4959,5007 lines on top of a broad H$\beta$ component, fixing the flux scalings using the [\ion{O}{iii}] doublet.  Note that we scale the H$\alpha$ and H$\beta$ profiles separately since the slit losses for the two wavelength regions may differ.   

Because the broad H$\beta$ fluxes in the Palomar and SDSS spectra are faint, we measure them differently.  Rather than fitting a Gaussian, which seems to overestimate the flux due to the poorly constrained width of the feature, we measure the broad flux by taking the integrated flux of the continuum-subtracted 4700--4950~\AA\ region after also subtracting out the narrow line fits.  For the SDSS spectrum, the error is the median relative flux error for the wavelength region.  For the \citet{ho95} Palomar spectrum, there is no reported error spectrum, though the estimated flux calibration errors are of order 30~per~cent. 

\begin{figure*}
\includegraphics[width=\linewidth]{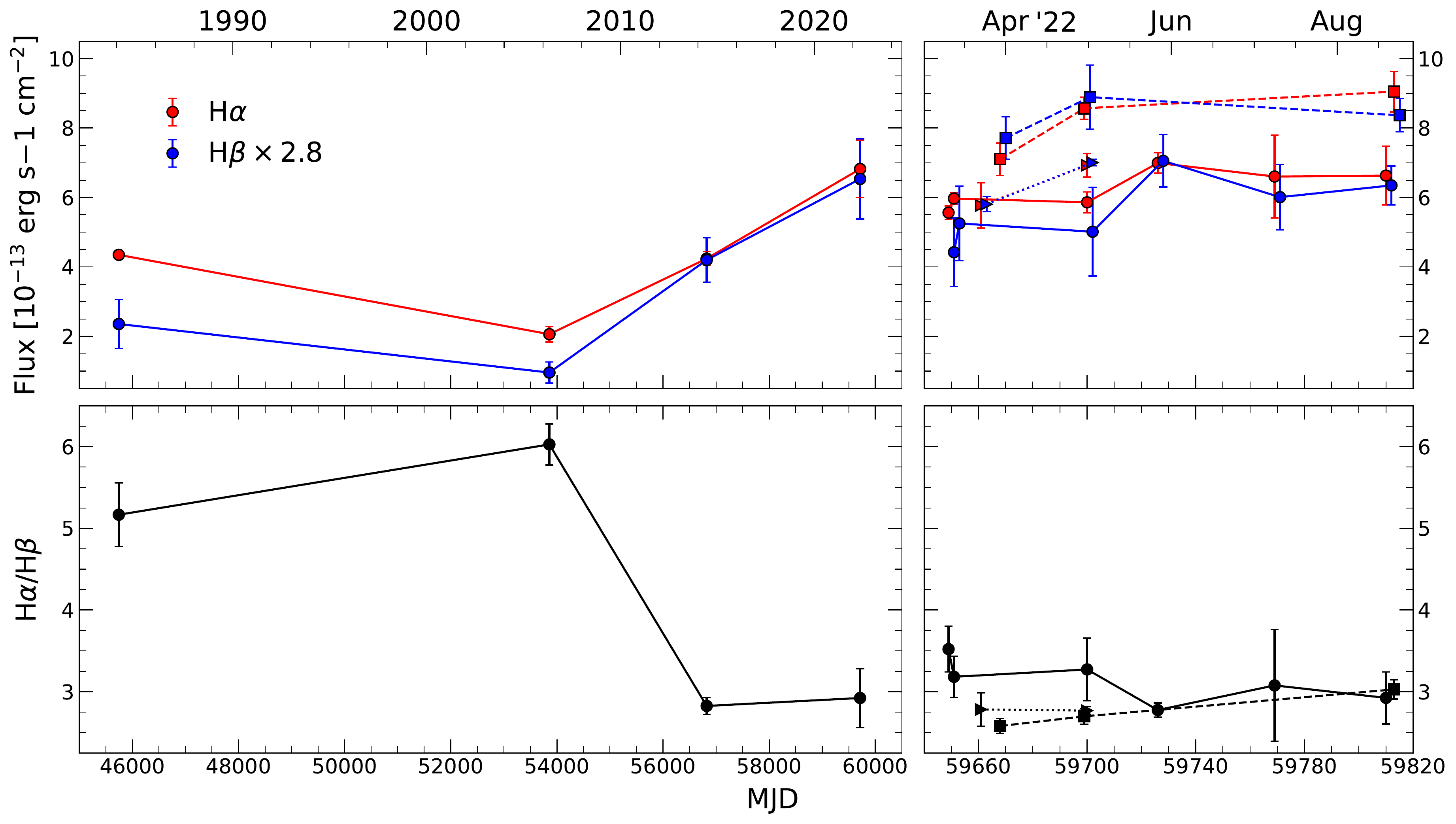}
\caption{\textbf{Top left:} Broad H$\alpha$ (red) and H$\beta$ (blue) fluxes calculated from the archival spectra, along with the mean Balmer fluxes from the 2022 data.  The H$\beta$ flux is scaled up by a factor of 2.8.  A significant increase in the H$\beta$ to H$\alpha$ flux ratio can be seen between 2006 and 2014 -- marking the change from a Type~1.8/1.9 to Type~1.  \textbf{Top right:} Broad Balmer fluxes for the 2022 spectra.  The three different marker shapes correspond to the different instruments (circles -- SNIFS; squares -- Keck; triangles -- MODS).  Since the fluxes calculated from roughly contemporaneous spectra (MJD~59700) differ, there are clearly systematic uncertainties.  \textbf{Bottom Left}: The H$\alpha$/H$\beta$ Balmer decrements of the archival spectra and the mean Balmer decrement of the 2022 spectra.  \textbf{Bottom right:} The Balmer decrements of the 2022 spectra.  Since 2014, the Balmer decrements are consistent with a single value of 2.6--3.0.}
\label{fig:balmer}
\end{figure*}

The final results are shown in Figure~\ref{fig:balmer} and listed in Table~\ref{tab:balmer}.  In addition to significant changes over the decades in the overall Balmer flux, we can see that the H$\alpha$/H$\beta$ Balmer decrement drops from a significantly reddened 5--6 in 1984/2006 to $\sim$2.8 in 2014, consistent with conventional case~B recombination and with typical Type~1 AGNs \citep{dong08}.  This change in the decrement is, effectively, the changing-look.  Furthermore, whereas the H$\alpha$ fluxes in the 1984 and 2014 spectra are roughly equivalent, the H$\beta$ flux in 2014 is a factor of two larger than in 1984.

We track the evolution of the Balmer flux in our 2022 spectra in Figure~\ref{fig:spec_full}.  We use the same methods to fit the H$\alpha$ and H$\beta$ profiles, scaling them to the SDSS spectrum.  The Balmer line fluxes and Balmer decrement are shown in Figure~\ref{fig:balmer} and listed in Table~\ref{tab:balmer}.  Despite our attempts at maintaining consistent flux scalings, there are discrepancies in the Balmer fluxes \textit{and} the decrement between the different instruments.  This is most likely due to different slits, apertures, and observing conditions.  In addition, the SNIFS dichroic overlap lies on the [\ion{O}{iii}] line complex and likely affects the flux scalings (and thus the H$\beta$ fluxes), making the Balmer decrements from the SNIFS spectra less accurate.  However, we stress that these differences between the observations in 2022 are still not as large as the differences between the fluxes and Balmer decrements from before 2014 (before the changing-look) and during 2022.

Looking only at the four contemporaneous MODS and Keck spectra (i.e., ignoring the last Keck spectrum), the overall fluxes are different by a factor of $\sim$1.3, but the Balmer decrements are the same, 2.6--2.8, given the errors.  This decrement is also, within errors, equivalent to the 2014 value despite the 2022 Balmer fluxes being higher than the 2014 fluxes.  In the third and last Keck spectrum, the Balmer decrement has increased slightly to $\sim$3.0.  This could be related to the flux drop around MJD~59810 (see Fig.~\ref{fig:swift-lc}), but this increase is not statistically significant (< 3$\sigma$, see Tab.~\ref{tab:balmer}). Overall, the Keck and MODS data indicate that the overall broad line flux is slightly increasing over time.  This trend can also be seen in the H$\alpha$ fluxes from the SNIFS spectra.  

The broad line widths do not significantly change over the course of the 2022 observations, with FWHMs of H$\alpha$ and H$\beta$ of 3600--3800~km~s$^{-1}$ and 4600--4800~km~s$^{-1}$, respectively.  The difference between the two widths is not uncommon, and is often attributed to the ``red shelf'' of H$\beta$ created by other emission lines \citep{derobertis85,marziani96,veron02}.  While there is no noticeable evolution of the H$\alpha$ FWHM in 2022, we measure the FWHMs of the broad H$\alpha$ in the Palomar 1984 and SDSS 2006 spectra to be $\sim$4100~km~s$^{-1}$.  This is moderately broader than the 2022 values and can be attributed to BLR ``breathing'' (e.g., \citealt{barth15,wang20}), a phenomenon where broad lines get narrower as luminosity increases, roughly following FWHM~$\propto L^{-1/4}$.  Based on this, the $\sim$10~per~cent change in FWHM corresponds to a $\sim$50~per~cent change in luminosity, which is, roughly, the change we see in H$\alpha$ flux between the Palomar~1984 and average 2022 values.  

In Figure~\ref{fig:spec_full}, we highlight the detection of the blue and red multiplets of \ion{Fe}{ii}, commonly seen in AGN spectra (e.g., \citealt{kovacevic10}).  These appear to be roughly constant with time.  Meanwhile, a broad emission feature centered around \ion{He}{ii}~$\lambda$4686 appears to slowly brighten over time.  This is most noticeable when comparing the MJD~59651 and MJD~59810 SNIFS spectra.  Other than this \ion{He}{ii} evolution and the broad Balmer flux changes, we do not observe any other significant changes in the optical spectra.

The changing-look phenomena in the optical lines is matched by similar changes in the NIR, as seen in Figure~\ref{fig:irspec}.  In the archival 2010 IRTF spectrum, there are no obvious broad emission lines.  We do identify several narrow emission lines, including [\ion{S}{ii}]~$\lambda$9071, [\ion{S}{iii}]~$\lambda$9533, and \ion{He}{i}~$\lambda$10833.  The lack of broad lines in the 2010 spectrum implies that NGC~5273 was yet to undergo the changing-look event that we know occurs by 2014 in the optical, and is consistent with the low UV fluxes during 2005--2011.  In the 2022 spectra, broad Paschen lines, including Pa$\beta$, Pa$\gamma$, Pa$\delta$, Pa$\epsilon$, and Pa9, are clearly present.  There are also broad \ion{O}{i}~$\lambda$8849, \ion{O}{i}~$\lambda$11290, and \ion{He}{ii}~$\lambda$10126 lines.  Additionally, the broad profile around Pa$\gamma$ is not centered on it, suggesting the presence of an overlapping broad \ion{He}{i}~$\lambda$10833 line.  We observe a broad Pa$\alpha$ line in the 2022 spectra, but due to the strong telluric features we are unable to reliably measure its properties.  The broad Pa$\beta$ line has a width of 3600--3900~km~s$^{-1}$, comparable to H$\alpha$ and narrower than H$\beta$.  These lines are all commonly seen in the NIR spectra of AGNs \citep{riffel06,landt08,lamperti17}.  

This is one of the few examples of a changing-look AGN in the NIR -- one other recent discovery being NGC~3786 \citep{son22}.  Some key differences between the changing-look events NGC~5273 and NGC~3876 are that the latter showed signs of heavy dust extinction, with relatively weak changes in the optical flux and broad emission lines compared to the NIR.  Furthermore, the broad emission lines in NGC~3876 are systematically redshifted by up to $\sim$900~km~s$^{-1}$, leading the authors to attribute the changing-look to be caused by a dust-obscured TDE.  There is no evidence that the changing-look in NGC~5273 is due to a TDE.  Furthermore, there is strong evidence against the changing-look being caused by changing obscuration or dust-reddening in the BLR -- namely, the lack of broad lines in the 2010 IRTF spectrum.  If obscuration or dust was causing the weak H$\beta$ and large Balmer decrement in the archival spectrum, one would expect to still see broad Paschen lines, as they would be less extincted than H$\alpha$.  Since we do not see these lines, changing obscuration is a poor explanation for the NIR spectral changes.


\section{X-ray Spectroscopy}\label{sec:xray}

Here we analyze the \swift, \xmm, \nicer, \chandra, and \suzaku\ X-ray data from 2000 to 2022 and try to connect the X-ray properties to the UV/optical/IR photometry and spectroscopy.

\begin{figure*}
\includegraphics[width=\linewidth]{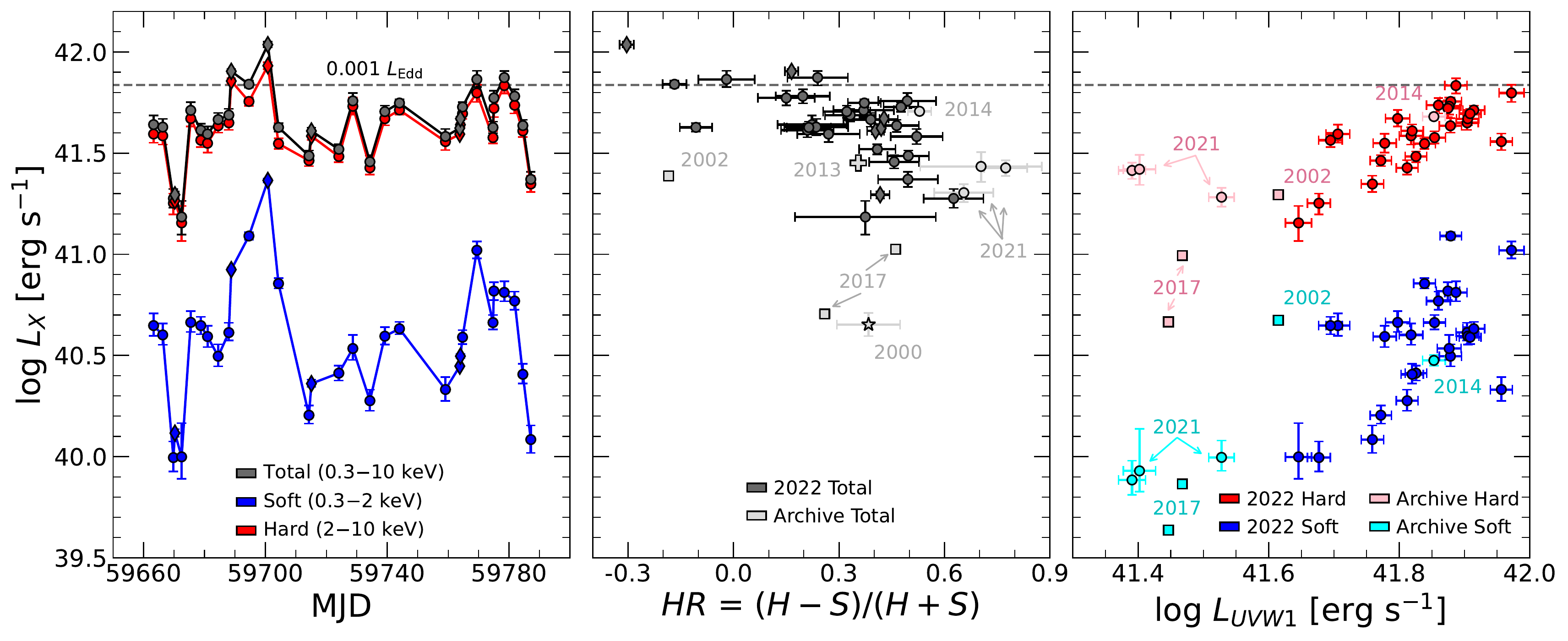}
\caption{\textbf{Left:} Hard (2--10~keV, red), soft (0.3--2~keV, blue), and total (0.3--10~keV, darkgray) X-ray luminosity lightcurves for the 2022 epochs.  Circles correspond to \swift\ observations, and diamonds correspond to \nicer\ observations.  \textbf{Center:} \hr\ plotted against total X-ray luminosity.  Darkgray and lightgray correspond to 2022 and archival data, respectively, and circles, diamonds, squares, star, and cross correspond to the \swift, \nicer, \xmm, \chandra, and \suzaku\ data, respectively. \textbf{Right:} \textit{UVW}1 luminosity $L_{\rm \textit{UVW}1}$ plotted against soft and hard luminosities.  Red/blue and pink/cyan correspond to 2022 and archival hard/soft flux, respectively.  Note that the archival \swift\ datapoint that ``fits in'' with the 2022 \swift\ datapoints is the one from 2014.}
\label{fig:xrays}
\end{figure*}

\subsection{Analysis of the flux}\label{sec:xflux}

Due to the relatively few counts in the \swift\ XRT and \chandra\ ACIS data, we started with a simple \textsc{xspec} model \textsc{tbabs $\times$ (zbbody + ztbabs $\times$ zpowerlw)} with which we can analyze all our data to calculate fluxes and hardness ratios.  The redshifts are fixed to the redshift of the source, $z=0.00342$.  The \textsc{zbbody} component is required to produce the observed ``soft excess'' that is common in AGN spectra \citep{walter93,gierlinski04,bianchi09-cat,hinkle21}, though the physical origin of these soft photons is unknown \citep{petrucci18,garcia19,petrucci20}. \textsc{tbabs} is meant to account for Galactic absorption and is fixed at the observed value of $N_{\rm H, Gal} = 9.2\ee{19} \rm ~cm^{-2}$ \citep{kawamuro16}, while \textsc{ztbabs} is meant to absorb the intrinsic power-law spectrum \textsc{zpowerlw} with column density $N_{\rm H}$.  Initially, we applied the \textsc{ztbabs} to the \textsc{zbbody} component as well, but this led to poor fits to the data.  We also attempted to model the soft excess without the \textsc{zbbody} component and instead replace \textsc{ztbabs} with a partial absorber \textsc{zpcfabs} and allow for soft flux to ``leak'' out from the power-law, but this also led to poor fits.  Due to the small count totals for much of the data, the \textsc{zbbody} temperature $kT$ and \textsc{zpowerlw} photon index $\Gamma$ are assumed to be constant between the different spectra, with best global fit values of $kT = 0.128 \pm 0.001 \rm ~keV$ and $\Gamma = 1.361 \pm 0.011$. The power-law slope is shallower than the previously reported values of 1.7--1.8 \citep{pahari17,vincentelli20}, but this is probably due to not including a reflected component (e.g., \textsc{xillver}, \citealt{dauser10,dauser13}) in our model, which would steepen $\Gamma$ to their values.  Nevertheless, our fitted values of $kT$ and $\Gamma$ are in line with those observed in other Type~1 and 1.8/1.9 AGNs \citep{gierlinski04,bianchi09-cat,ricci17,hinkle21}. We vary the \textsc{zbbody} and \textsc{zpowerlw} normalizations and the \textsc{ztbabs} $N_{\rm H}$ for each epoch.  

From this model, we calculate the soft (0.3--2~keV), hard (2--10~keV), and total (0.3--10~keV) fluxes.  These values are uncorrected for Galactic and host galaxy absorption.  We use the model photon counts to calculate the hardness ratio $\textit{HR} =(H-S)/(H+S)$, where $H$ and $S$ are the hard and soft photon counts, respectively.  By using model \textit{photon} counts instead of \textit{detector} counts, we are able to accurately compare between different instruments with different spectral responsivities.  The fluxes, \textit{HR}, and $N_{\rm H}$ are listed in Table~\ref{tab:xflux}.  Figure~\ref{fig:xrays} shows the evolution of the hard, soft, and total luminosities and the dependence of the X-ray flux on the \textit{HR} and the UV flux.  In the left panel, we can see that the soft and hard fluxes are both varying and that the soft flux varies more than the hard flux. In the center panel, the \hr\ varies from $-$0.3 to 0.6 over the 2022 observations with a trend of being ``softer-when-brighter'', meaning that the \hr\ increases as total X-ray flux decreases, though this trend seems to only hold for the 2022 data.  This panel also includes the archival \swift, \xmm, \chandra, and \suzaku\ measurements.  While the 2013 \suzaku\ and 2014 and 2021 \swift\ data seem to lie in the same parameter range as the 2022 \swift\ and \nicer\ data, the 2000 \chandra\ data and the 2002 and 2017 \xmm\ data do not.  This panel also emphasizes the large changes in X-ray flux between the 2000 \chandra\ and 2002 \xmm\ observations, separated in time by 649~d, as well as the change between the two 2017 \xmm\ observations, separated in time by only 2~d.

In the right panel of Figure~\ref{fig:xrays}, we show the soft and hard X-rays with the simultaneously obtained \swift~UVOT and \textit{XMM}-OM \textit{UVW}1 monochromatic luminosity $L_{\rm \textit{UVW}1}$.  The most noticeable trend is that the soft X-ray flux correlates well with the UV flux, while the hard X-ray flux is more stochastic below $L_{\rm \textit{UVW}1} \sim 10^{41.6}$~erg~s$^{-1}$.  For example, the 2017 \xmm\ data show a decreasing hard X-ray flux with a decreasing UV flux, but the 2021 \swift\ data show a flat (or perhaps even increasing) hard X-ray flux with a decreasing UV flux.  The 2002 \xmm\ data stand out due to the anomalously bright soft X-ray flux given the UV flux.

\begin{figure*}
\includegraphics[width=0.95\linewidth]{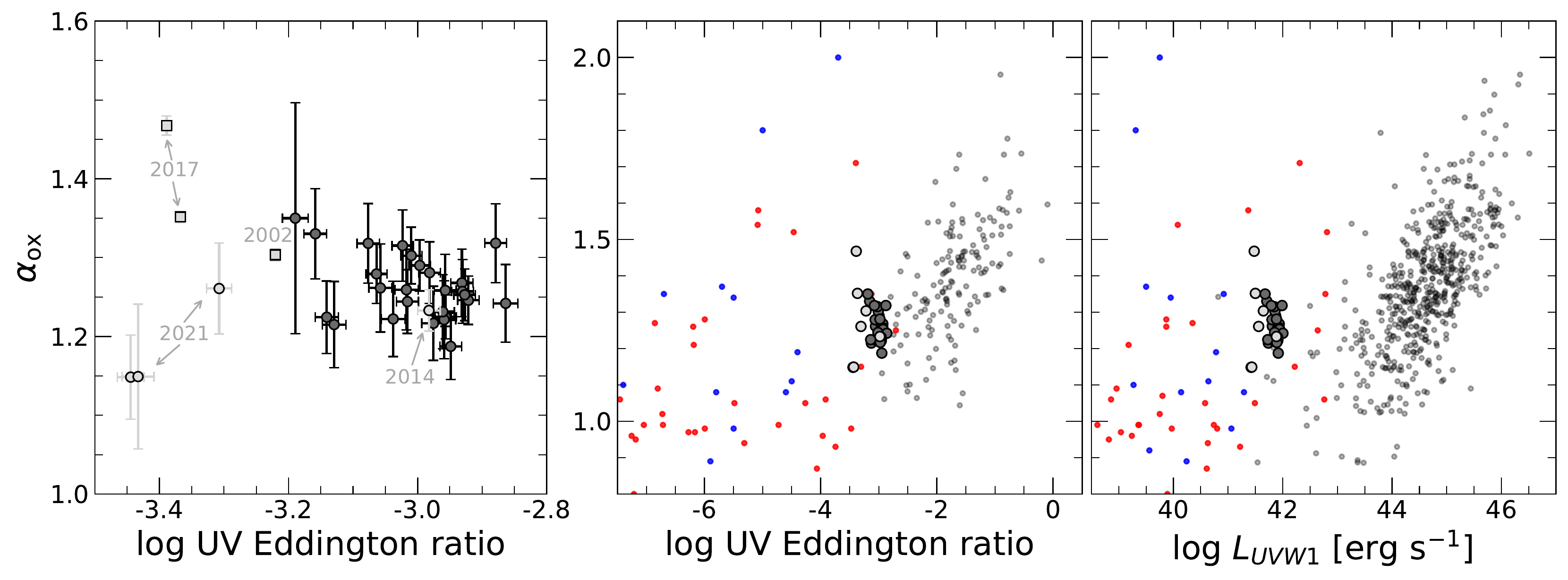}
\caption{\textbf{Left:} Spectral index $\alpha_{\rm ox}$ as a function of UV Eddington ratio ($L_{\textit{UVW}1}/L_{\rm Edd}$).  Colors and symbols are the same as in Figure~\ref{fig:xrays}. As seen in Fig.~\ref{fig:xrays}, the archival 2014 \swift\ data ``fit in'' with the 2022 \swift\ data, whereas most of the other archival data do not.  The 2002 \xmm\ data appear to approach the parameter region of the 2014/2022 \swift\ data. \textbf{Center:} Same as left, with data for typical AGNs (black, \citealt{lusso10}) and LLAGNs (blue, \citealt{maoz07}; red, \citealt{xu11}).  Note that the range of the axes are larger, particularly the luminosity axis. \textbf{Right:} Same as center, but using $L_{\textit{UVW}1}$ luminosity instead of the UV Eddington ratio.  There are more objects shown in the right panel because many objects lack a reliable $M_{\rm BH}$ measurement.}
\label{fig:aox}
\end{figure*}

We can also compare the UV and X-ray fluxes using the spectral index \citep{tananbaum79} 
\begin{equation}
\alpha_{\rm ox} = -0.384 \log{(L_{\rm 2keV}/L_{\rm 2500\mathring{A}})} ~.
\end{equation} 
$L_{\rm 2keV}$ is calculated directly from the \textsc{xspec} fits, while $L_{\rm 2500\mathring{A}}$ is equated with \textit{UVW}1 luminosity $L_{\textit{UVW}1}$ since the \textit{UVW}1 effective wavelength is 2600~\AA. In Figure~\ref{fig:aox}, we show $\alpha_{\rm ox}$ compared to the UV Eddington ratio, defined as $L_{\textit{UVW}1} / L_{\rm Edd}$, with $L_{\rm Edd} = 10^{44.8} \rm ~erg~s^{-1}$ based on the \citet{bentz14} $M_{\rm BH}$ measurement.  The $\alpha_{\rm ox}$ values are also given in Table~\ref{tab:xflux}.  In the left panel, we only show the evolution of NGC~5273, whereas in the center panel, we show this evolution with respect to comparison samples of ``typical'' AGNs \citep{lusso10} and LLAGNs \citep{maoz07,xu11}.  The archival 2014 \swift\ data lie in the same region as with the 2022 \swift\ data, whereas most of the other archival data do not.  The 2021 \swift\ data have shallower $\alpha_{\rm ox}$ (i.e., brighter X-rays relative to their UV flux), and the 2017 \xmm\ data have steeper indices.  Compared to other archival data, the 2002 \xmm\ data are more similar to the 2014 and 2022 \swift\ data.  The evolution of NGC~5273 in the center panel shows that it is closer to the region of UV Eddington ratio space occupied by the LLAGNs in 2017 and 2021, whereas it is closer to typical AGNs in 2014 and 2022 (and arguably 2002).  However, even when it overlaps in parameter space with typical AGNs, it does not seem to follow the positive correlation between UV luminosity and $\alpha_{\rm ox}$ that has been observed in typical AGNs (e.g., \citealt{lusso10}).  This correlation, and NGC~5273 deviating from it, is most prominent in the right panel of Figure~\ref{fig:aox}, where we show \textit{UVW}1 luminosity instead of UV Eddington ratio.  This panel also highlights how underluminous NGC~5273 is compared to the typical AGN in these studies, although much of the difference in the right panel is driven by NGC~5273 having a smaller SMBH mass than the typical AGN in \citet{lusso10}.

\begin{figure}
\includegraphics[width=0.95\linewidth]{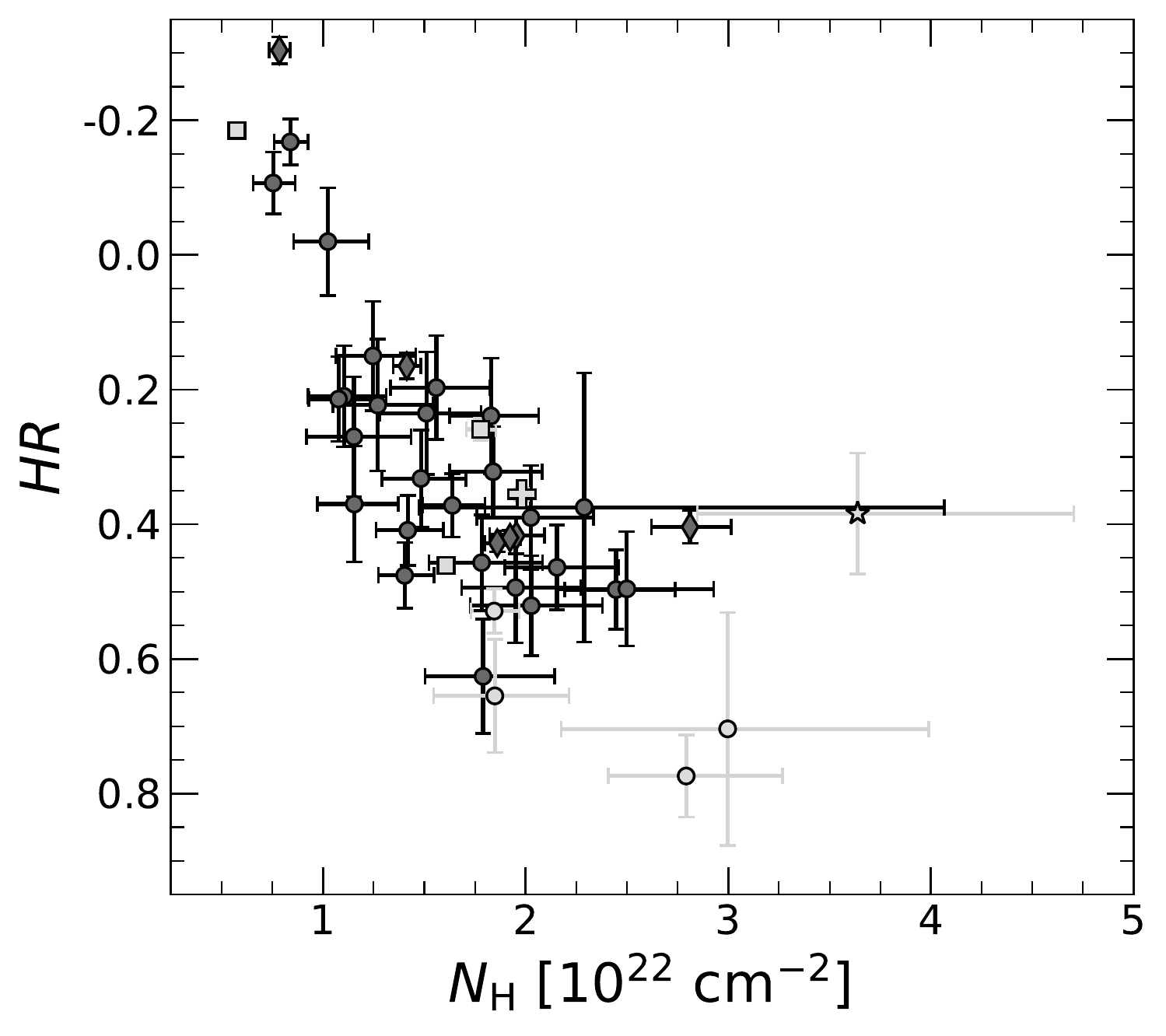}
\caption{\hr\ plotted against $N_{\rm H}$ using the blackbody+absorbed power-law model. Colors and symbols are the same as in Figure~\ref{fig:xrays}. The strong relation between the two implies that variable absorption is driving the changes in \hr, but it does not appear to be driving the changing-look phenomena (see Sec.~\ref{sec:xipcf}).}
\label{fig:nh-hr}
\end{figure}

We see in Figure~\ref{fig:nh-hr} that the \hr\ is clearly related to the absorbing column $N_{\rm H}$ of the power-law (\textsc{ztbabs}).   In the model, the soft photons come from the unabsorbed blackbody component, and the absorption only affects the hard photons coming from the power-law component.  The relation in Figure~\ref{fig:nh-hr} thus implies that the unabsorbed blackbody/soft excess becomes brighter as the power-law becomes less absorbed, in spite of the model allowing them to be independent of each other.  The easiest physical explanation is that the soft excess in NGC~5273 is not an independent component, but instead the result of variable absorption of the power-law component \citep{gierlinski04,turner09,cwang22}.

This implies we need a more complex model, but we can only apply a more complex model to the higher count \xmm\ and \nicer\ observations.  We considered a range of models with the general form of \textsc{tbabs $\times$ ABSORPTION $\times$ (POWERLAW + REFLECTION)}, where \textsc{tbabs} is always the Galactic absorption.  The main reason we replace a simple power-law with a power-law \textit{and} reflection is to keep $\Gamma$ in a reasonable range (>1.2).  All of our models included a 140~keV cutoff energy on the power-law based on the results of \citet{pahari17}, though this will not affect our results as we only consider energies below 10~keV.  We initially tried the model in \citet{vincentelli20}, with a neutral partial absorber (\textsc{zpcfabs}), a simple power-law (\textsc{zpowerlw}) and a reflection component modeled with \textsc{xillver} \citep{dauser10,dauser13}.  We tried a number of fits, allowing the absorbing column $N_{\rm H}$ and partial covering fraction $f_c$ of \textsc{zpcfabs} and the ionization parameter of the \textsc{xillver} reflector $\xi$ to vary between observations and also fitting them as single values between the spectra, leading to a series of $\chi^2/\nu$ values between 1.33--1.48.  Our fits were immediately improved by replacing the neutral partial absorber \textsc{zpcfabs} with a ionized partial absorber \textsc{zxipcf} \citep{reeves08}.  Again, we varied $N_{\rm H}$, $f_c$, and the $\xi$ values of \textsc{zxipcf} and \textsc{xillver} being fit as single values between the different spectra, and additionally attempted fitting the two ionizing parameters $\xi$ from \textsc{xillver} and \textsc{zxifpcf} as a single value, leading to a range of $\chi^2/\nu$ values between 1.17--1.26. 

\begin{figure*}
\includegraphics[width=\linewidth]{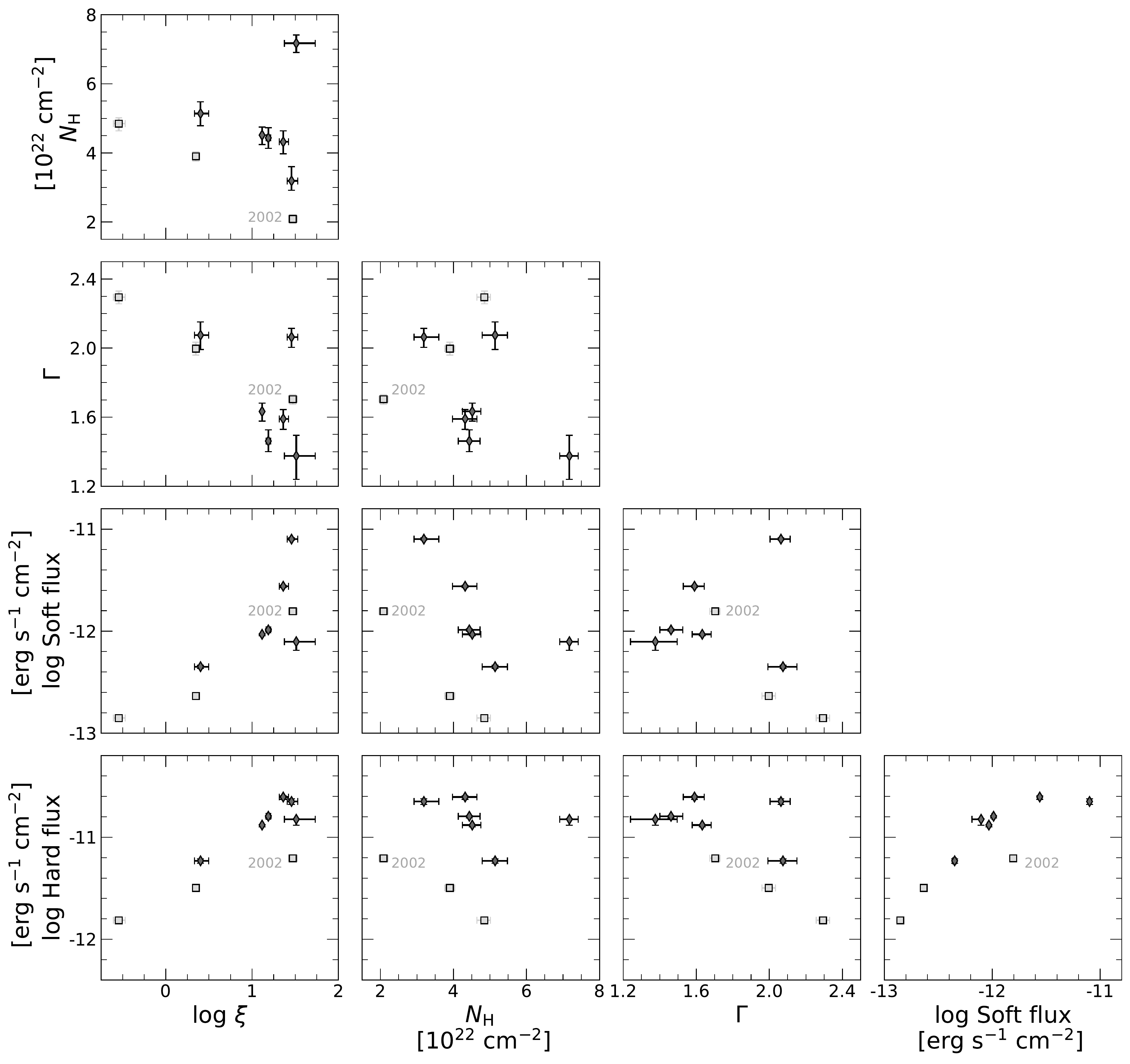}
\caption{Model parameters $\xi$, $\Gamma$, and $N_{\rm H}$ and the hard and soft fluxes for the ionized absorber model. Colors and symbols are the same as in Figure~\ref{fig:xrays}. In each panel, the 2002 \xmm\ data are labeled, and so the other two archival datapoints are the 2017 \xmm\ data.  The only noticeable trends are an inverse correlation between $\xi$ and $\Gamma$ as well as a positive correlation between $\xi$ and soft flux. }
\label{fig:zxipcf}
\end{figure*}

Adding an additional constant neutral absorber component \textsc{ztbabs} marginally improved the fit, with $\Delta \chi^2/\nu \approx 0.05$.  We also attempted other \textsc{POWERLAW + REFLECTION} trials with \textsc{pexrav} \citep{magdziarz95} and \textsc{pexmon} \citep{nandra07}. For these models, the main variable component is the scaling of the reflection component $f_{\rm refl}$.  We allowed it to vary between the spectra and also fit it with a single value, and find that there are only marginal fit improvements with $\Delta \chi^2/\nu \approx 0.03$.  Overall, the goodness of fit for all the attempted models spanned $1.19 \leq \chi^2/\nu \leq 1.24$.  Despite the differences between all the various models, all of the observed trends between fitted parameters -- namely, $N_{\rm H}$, $\xi$, and $\Gamma$ -- are roughly the same as those we now discuss in detail.

\subsection{Ionized absorption model}\label{sec:xipcf}

Based on these experiments, we settle on using our final model \textsc{tbabs $\times$ ztbabs $\times$ zxipcf $\times$ pexmon}.  As before, the \textsc{tbabs} component corresponds to Galactic absorption, while the \textsc{ztbabs} corresponds to neutral absorption occurring within NGC~5273.  We initially allow the neutral absorption to vary but find that the values are all consistent with a common value of $N_{\rm H, \textsc{ztbabs}} =(3.95 \pm 0.30)\ee{20} \rm ~cm^{-2}$.  Similarly, for the \textsc{zxipcf} component, we initially vary the covering fraction $f_{\rm c}$ but find it is roughly consistent with a common value of $f_{\rm c} = 0.966 \pm 0.007$, which is also consistent with the value found in \citet{vincentelli20}.  We also allow $f_{\rm refl}$ in \textsc{pexmon} to vary, and again we find it is consistent with a common value of $f_{\rm refl} = 1.108 \pm 0.90$, roughly equivalent to isotropic reflection.  We use an inclination of $i = 40\degr$ to match previous models \citep{pahari17,vincentelli20}.  We keep the metal and Fe abundances $A$ and $A_{\rm Fe}$ fixed at the default values of 1.  This allowed us to vary the remaining parameters of the model -- $N_{\rm H}$, $\Gamma$, and $\xi$ -- epoch to epoch.  We show the resulting parameters of our final model for the 9 spectra in Figure~\ref{fig:zxipcf} and in Table~\ref{tab:xparams} along with the soft and hard fluxes.  One indication that our final model is physically valid is that we see a correlation between $\xi$ and $\Gamma$ -- as $\Gamma$ decreases and the power-law gets harder, $\xi$ increases and the absorber becomes more ionized.  Additionally, we see that the soft flux increases as $\xi$ increases -- this can be understood as the absorber letting more soft flux escape as the absorber becomes more ionized.  

Notably, we do not see any obvious relations between $N_{\rm H}$ and the other parameters or fluxes, and we do not see coherent changes in $N_{\rm H}$ over time.  This seems to indicate that the variable $N_{\rm H}$ is unrelated to the prominent trends in X-ray and UV luminosity that we discuss in Section~\ref{sec:xflux}, and so it is reasonable to conclude that the variable $N_{\rm H}$ is unrelated to the changing-look phenomenon in NGC~5273.  This is in line with other studies of changing-look events, which often find no connection between the changes in the broad emission lines and those in $N_{\rm H}$ (e.g., \citealt{shappee14,lamassa15,guolo21}). This also highlights the distinction between changing-state events and the variable absorption changing-obscuration events discussed in Section~\ref{sec:intro}.


\section{Discussion} \label{sec:discuss}

\begin{figure*}
\includegraphics[width=0.95\linewidth]{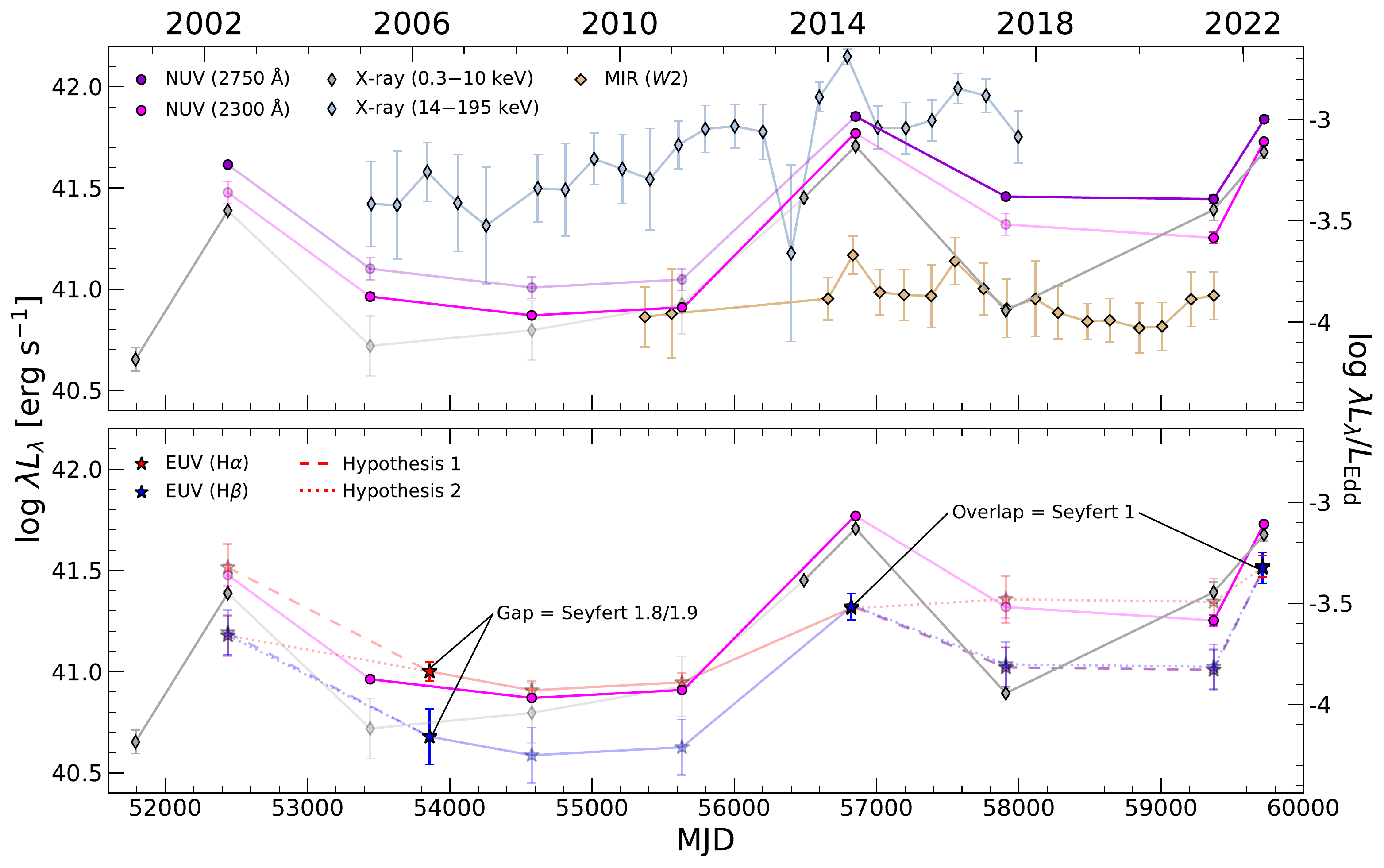}
\caption{\textbf{Top:} Evolution of the X-ray, UV, and MIR luminosities over time.  0.3--10~keV X-ray luminosity is from the various X-ray observations (see Sec.~\ref{sec:xflux}), and the 14--195~keV X-ray luminosity is from the \swift~BAT lightcurve (see Fig.~\ref{fig:bat}).  NUV (2750~\AA) luminosity is calculated using from \swift~UVOT and \textit{XMM}-OM \textit{UVW}1 observations, and NUV (2300~\AA) using \swift~UVOT \textit{UVM}2 and GALEX \textit{NUV} (see Sec.~\ref{sec:phot}).  Opaque points are actual observations, and transparent points are interpolations based on observed NUV and X-ray colors. \textbf{Bottom:} Evolution of the X-ray, NUV, and EUV luminosities over time.  EUV (H$\alpha$) and EUV (H$\beta$) luminosities are calculated based on the broad H$\alpha$ and H$\beta$ fluxes, respectively, assuming case~B recombination.  During the epochs where the two EUV luminosities match, it is a Type~1, and it is a Type~1.8/1.9 where the two do not match.  As before, transparent lines are interpolations, but with the EUV there are two hypotheses: (1 - dashed) NGC~5273 undergoes only one changing-look event between 2011 and 2014; or (2 - dotted) NGC~5273 undergoes multiple changing-look events from 2000 to 2022.  Both use the observed ratios of EUV to NUV (2300~\AA) fluxes from when NGC~5273 is a confirmed Type~1.8/1.9 (2005--2011) and Type~1 (2014/2022).}
\label{fig:lum_lc}
\end{figure*}

We show the long-term evolution of the IR, UV, and X-ray luminosity in the top panel of Figure~\ref{fig:lum_lc}.  Following Section~\ref{sec:observations}, we interpolate between observations with different near-UV (NUV) filters to get the ``full'' lightcurves at 2300~\AA\ (\textit{NUV} + \textit{UVM}2) and 2750~\AA\ (\textit{UVW}1).  Similarly, we estimate the 0.3--10~keV X-ray flux from 2005--2011 using the contemporaneous 14--195~keV \swift~BAT fluxes and observed colors (i.e., ratio between 0.3--10~keV and 14--195~keV fluxes) of the 2014 and 2017 \swift~XRT and \xmm\ observations. In the bottom panel of Figure~\ref{fig:lum_lc}, we show the extreme-UV (EUV, <912~\AA) luminosity implied by the H$\alpha$ and H$\beta$ fluxes and case~B recombination.

Here we present a plausible reconstruction of the evolution of NGC~5273 from 2000 to 2022.  From 2006 until 2011 at the earliest, NGC~5273 is a Type~1.8/1.9 Seyfert (see Fig.~\ref{fig:spec_archive}).  Between 2011 and 2014, NGC~5273 changes-look to become a Type~1 with the emergence of broad H$\beta$, H$\gamma$, and other broad Balmer lines.  It is likely that the broad Paschen lines also emerge around this time (see Fig.~\ref{fig:irspec}).  This changing-look event is coincident with an increase in brightness at all wavelengths (see Fig.~\ref{fig:lum_lc}).  It peaks and then decreases in brightness until mid-2016, when a second flare occurs (see Figs.~\ref{fig:asassn-lc} and \ref{fig:lum_lc}).  After the flare, it fades, and NGC~5273 enters a state with lower flux than in 2014, but with a slightly higher UV and 14--195~keV X-ray flux than in 2005--2011.  Beginning in 2021, NGC~5273 begins to slowly brighten again.  Compared to 2017, the 0.3--10~keV X-ray flux increases, but the UV remains roughly constant.  The 2021 X-ray fluxes are comparable to the 2013 fluxes.  In early-2022, NGC~5273 is a Type~1 Seyfert (see Figs.~\ref{fig:spec_archive} and \ref{fig:spec_full}).  While the Balmer fluxes are in 2022 are overall higher than in 2014, the Balmer decrement is roughly the same (see Fig.~\ref{fig:balmer}).  Throughout 2022, it continues to vary, with evidence in the X-rays for varying ionized absorption (see Figs.~\ref{fig:nh-hr} and \ref{fig:zxipcf}).  These variations are not noticeably different from those observed in 2017 and 2021, but both the hard and soft fluxes are on average higher (see Fig.~\ref{fig:xrays}).  The X-rays in 2014 and 2022 are comparable both in terms of flux and \textit{HR}.  The lack of NIR broad lines in the 2010 archival spectrum (see Sec.~\ref{sec:spec}), as well as the lack of coherent trends in $N_{\rm H}$ variability (see Sec.~\ref{sec:xipcf}), indicate that the changing-look event was not due to variable obscuration -- this is not a changing-obscuration event (see Sec.~\ref{sec:intro}).

Before 2006, the evolution of NGC~5273 is complicated. It is assumed that NGC~5273 was a Type~1.8/1.9 in 2000 based on the 1984 Palomar spectrum and the other observations before 2000 (see Sec.~\ref{sec:intro}). The X-rays in 2000 are much more similar to the X-rays in 2017 than to any other epoch, both in terms of flux and \hr\ (see Fig.~\ref{fig:xrays}).  Meanwhile, the hard X-ray and UV fluxes in 2002 are marginally fainter than most of the fluxes in 2014 and 2022 data, but the soft X-ray fluxes are similar, making it  much softer in 2002 than in most of 2022. The spectral index $\alpha_{\rm ox}$ in 2002 is also similar to $\alpha_{\rm ox}$ in 2014 and 2022 (see Fig.~\ref{fig:aox}).  Overall, the data indicate that there is a large X-ray flare in 2002 and that the UV and X-ray properties in 2002 and 2022 are similar.  It is unclear what happens between 2002 and 2013, as there are no reported 0.3--10~keV X-ray data from this period, though the 14--195~keV \swift~BAT data (see Fig.~\ref{fig:lum_lc}) suggest that the X-ray flux slowly increased without any significant flares.

The interpretations of the data from 2000 to 2022 can be split into two hypotheses: (1) NGC~5273 undergoes only one changing-look event between 2011 and 2014, going from a Type~1.8/1.9 to a Type~1; or (2) NGC~5273 undergoes multiple changing-look events.  The second hypothesis posits that NGC~5273 spends most of the 2000--2022 period as a Type~1.8/1.9, but temporarily changes-look to a Type~1 in 2002 and again in 2014, reverting back to a Type~1.8/1.9 by 2005 and 2017, respectively.  In 2022, it again becomes a Type~1.  The idea that 2002 and 2014 are temporary changing-look events is based on the relative brightness in 2002 and 2014 and the relative faintness in 2000, 2005--2011, and 2017.  Since it is brighter in 2014 and 2022 when it confirmed to be a Type~1 and fainter in 2006--2011 when it is a confirmed Type~1.8/1.9, it is not unreasonable to assume that significant changes in brightness correlate with changing-look events.  This hypothesis, as well as the alternative, are presented in the bottom panel of Figure~\ref{fig:lum_lc}.  One key difference is that the first hypothesis predicts higher EUV (i.e., H$\alpha$) flux in 2002 than the second, while the first predicts lower EUV flux from 2017--2021 than the second.  These are based on the relative NUV to EUV ratios from when NGC~5273 is a confirmed Type~1.8/1.9 or Type~1. 

We favor the second hypothesis, and this would not be the first case of an AGN repeatedly changing-look. NGC~1566 \citep{alloin86,dai18,oknyansky19,parker19}, NGC~2992 \citep{gilli00,trippe08,guolo21}, NGC~4151 \citep{czerny03,shapovalova08,guo14,oknyanskij16}, and NGC~5548 \citep{iijima92,czerny99,bon16,mathur17,kovacevic18} are well-studied examples of AGNs that have repeatedly flared and changed-look over decades of observation.  These repeating changing-look events are spaced out over 5--10~yr, depending on the AGN, which is comparable to the observed gaps in UV brightness between 2002--2014 and 2017--2022 in NGC~5273.  This would also explain the variable typing of the AGN in previous decades (see Sec.~\ref{sec:intro}). Indeed, theoretical explanations for changing-look behavior, such as ``unstable'' transition zones between radiatively efficient and inefficient parts of the disc, could produce multiple flares on timescales of years to decades \citep{sniegowska20}.  NGC~5273 could prove to be an interesting test case of the changing-look phenomenon.  The 2022 changing-look event appears to be distinct from the 2002 and 2014 flares in that it is brighter in the UV, has stronger broad Balmer lines, and has a slow rise time (see Figs.~\ref{fig:asassn-lc} and \ref{fig:lum_lc}).  There is also a slow increase in the overall 14--195~keV X-ray flux that could indicate long-term changes to the accretion rate of NGC~5273.  It can also be inferred that whatever change NGC~5273 is undergoing, it is being driven by the innermost accretion flow, as the 2021 data show a significant increase in X-ray flux (compared to 2017) that is not matched in the UV (see Fig.~\ref{fig:lum_lc}).  Rather than a short flare, the 2022 event could be a more ``stable'' change to the accretion rate, though continued monitoring of the source will show whether this the case or not.

While we prefer this hypothesis, we cannot prove it without spectra from 2002 or 2017--2021, nor can we prove the alternative, where NGC~5273 changes-look only once.  In this scenario, before 2014, it is always a Type~1.8/1.9, even during the flare in 2002, and after 2014, it is always a Type~1, even during the low flux state from 2017--2021.  The main argument in favor of this hypothesis is that large changes in flux are not always coincident with changing-look events.  \citet{macleod16} found that at least 15~per~cent, but not all, of strongly variable quasars underwent changing-look events over 8--10~yr rest-frame timescales.  While the hypothesis has interesting implications for what is considered ``normal'' AGN variability, it poses a challenge for future searches if changing-look events can be uncorrelated with UV or X-ray variability and can only be identified using spectra.  It is also at odds with recent claims that connect changing-look behavior with changes in UV and/or X-ray luminosity, and thus, the Eddington ratio $\lambda_{\rm Edd}$ (e.g., \citealt{guolo21,jwang22}). If spectra taken during these periods of missing data exist, perhaps in some unpublished archive, the authors would appreciate such spectra being made available.   

Regardless of the number of changing-look events in NGC~5273, we can track the evolution of $\lambda_{\rm Edd}$ before and after the confirmed changing-look.  We can compute $\lambda_{\rm Edd}$ by converting the 2--10~keV luminosity $L_{\rm X}$ into a bolometric luminosity $L_{\rm bol}$.  For AGNs of this luminosity, the bolometric correction $K_X = L_{\rm bol}/L_{\rm X}$ ranges from 6--20 \citep{ho08,vasudevan09,duras20}.  Adopting a nominal $K_X = 15$, we find the mean Eddington ratio for the 2022 is $\log \lambda_{\rm Edd} = -1.97 \pm 0.03$, with the caveat that uncertainties in $M_{\rm BH}$ and the spread in $K_X$ values introduce systematic errors of 0.15~dex and 0.41~dex, respectively.  This is comparable to the value of $-2.05$ calculated by \citet{pahari17} using the 2014 \swift~XRT and \nustar\ data after adjusting for a different $K_X$ used in their paper.  The low flux states in 2000 and 2017 yields $\log \lambda_{\rm Edd}$ of $-3.01 \pm 0.06$ and $-2.77 \pm 0.01$, respectively, highlighting how significantly $\lambda_{\rm Edd}$ changes. 

One can also compare the changes in $\lambda_{\rm Edd}$ to the changes in H$\alpha$ and H$\beta$ flux.  Whereas the changes between 2000/2017 and 2022 in $\lambda_{\rm Edd}$ range between 0.8--1.0~dex, the changes between the 2006 and 2022 in the H$\alpha$ and H$\beta$ fluxes are 0.52~dex and 0.84~dex, respectively.  The offset between the scale of the $\lambda_{\rm Edd}$ changes and of the Balmer flux changes can be attributed to the intrinsic Baldwin effect \citep{baldwin77,pogge77}, where the equivalent width ($EW$) of emission lines decreases as the continuum luminosity $L_{\rm cont}$ increases, or $EW \propto L_{\rm cont}^{\beta} \sim \lambda_{\rm Edd}^\beta$ with $\beta < 0$.  The closer $\beta$ is to 0, the more responsive the line is to continuum changes.  This can be rewritten in terms of an emission line luminosity $L_{\rm line}$ as 
\begin{equation}
\Delta \log L_{\rm line} = (\beta+1) \Delta \log \lambda_{\rm Edd} ~. 
\end{equation}
For conventional AGN variability (i.e., not changing-looks), H$\alpha$ and H$\beta$ show $\beta < -0.4$ \citep{rakic17}.  For NGC~5273, the changing-look leads to $\beta \sim -0.4$ for H$\alpha$  but $\beta \sim -0.1$ for H$\beta$.  While the H$\alpha$ emission remains as responsive to $\lambda_{\rm Edd}$ changes as in normal AGNs, the H$\beta$ emission is much more responsive to $\lambda_{\rm Edd}$ changes than in normal AGNs.  This extra-responsivity, along with the changing Balmer decrement and emergent NIR broad lines, shows that the changing-look event is linked to a change in how the BLR reprocesses the continuum emission, not just the BLR becoming brighter. 

Changes in the reprocessing mechanisms of the BLR could be due to changes in the ionization state of the BLR, as the ionizing radiation ought to change with increasing $\lambda_{\rm Edd}$.  Studies of LLAGNs, true Type~2s, and changing-looks have found that $\log \lambda_{\rm Edd}$ values around --3 to --2 may be a ``threshold'' $\lambda_{\rm Edd}$ below/above which broad emission lines disappear/appear (e.g., \citealt{bianchi12,macleod19,ruan19,guolo21,jana21,lyu22}). It has also been theorized that $\log \lambda_{\rm Edd} \sim -2$ is the transition from the standard radiatively-efficient thin disc \citep{shakura73} into the radiatively-inefficient/advection-dominated accretion flow (RIAF/ADAF, \citealt{ho08,xie12,yuan14}), implying a connection between the the accretion state of the AGN and its BLR.  Above, we estimate $\lambda_{\rm Edd}$ values similar to these as NGC~5273 transitions from Type~1.8/1.9 to Type~1, meaning that the changing-look events in NGC~5273 could very well be due to repeated crossings over this threshold $\lambda_{\rm Edd}$ value and transitions between accretion states.

Alternatively, or perhaps concurrently, the reprocessing changes could be due to changes in the BLR structure.  Alternative theories to the standard unified model have posited that BLR clouds are produced by a disc wind whose strength and properties depends on $\lambda_{\rm Edd}$ (e.g, \citealt{nicastro00,elitzur09,elitzur14,giustini19}).  Specifically, \citet{elitzur14} found that Type~1s, 1.2/1.5s, and 1.8/1.9s could be separated into different populations based on the quantity $L_{\rm bol} / M_{\rm BH}^{2/3}$, which on average decreases as one moves from Type~1 to 1.8/1.9.  For NGC~5273 in its state as a Type~1 in 2022 and a Type~1.8/1.9 in 2000, we calculate $\log \big(L_{\rm bol}/{\rm erg~s^{-1}}) -2/3 \log (M/\msun)$ to be 38.4 and 37.4, respectively, which are roughly in line with the findings of \citet{elitzur14}.  Changes in the wind due to changes in $\lambda_{\rm Edd}$ (and similar quantities) could thus lead to changes in the BLR, though it is unclear how this specifically affects the Balmer decrement or the NIR broad lines.  More studies of changing-look AGNs, especially those that change to/from intermediate Type~1.8/1.9s, are necessary to understand how the BLR evolves with changing $\lambda_{\rm Edd}$.


\section{Conclusions}\label{sec:conc}

We characterize the evolution of NGC~5273, an optical and NIR changing-look AGN.  We summarize our results below:
\begin{itemize}
\item[$\bullet$] The AGN varies by factors of 2 to 10 in IR to X-ray flux, with short flares in 2002, 2014, 2016, and a long ongoing flare beginning in late 2021. \\
\item[$\bullet$] NGC~5273 undergoes at least one changing-look event from Type~1.8/1.9 to Type~1 where broad optical and NIR emission lines appear in the spectra.  It is one of the few AGNs known to change-look in the NIR.\\
\item[$\bullet$] Because the changing-look is coincident in time with the 2014 flare, and with evidence for historical variability prior to 2000, it is likely that the other flares were temporary changing-look events from a baseline Type~1.8/1.9 into Type~1.\\
\item[$\bullet$] The changing-look phenomena in NGC~5273 is likely due to changes in how the BLR reprocesses the continuum emission as evidenced by the changing $\lambda_{\rm Edd}$, Balmer decrement, and emergent broad NIR emission lines.
\end{itemize}
Future studies of NGC~5273 and other changing-look AGNs will help us better understand the changing-look phenomenon, AGN structure, and the dynamics of accretion processes.   


\section{Acknowledgements}

We thank the referee for their helpful comments and suggested edits. 

We thank Prof. Misty C. Bentz (Georgia State University) for providing an optical spectrum from \citet{bentz14} and for helpful comments. 

We thank the \swift~PI, the Observation Duty Scientists, and the science planners for promptly approving and executing our \swift\ observations.

We thank Las Cumbres Observatory and its staff for their continued support of ASAS-SN. ASAS-SN is funded in part by the Gordon and Betty Moore Foundation through grants GBMF5490 and GBMF10501 to the Ohio State University, and also funded in part by the Alfred P. Sloan Foundation grant G-2021-14192.

J.N., C.S.K. and K.Z.S. are supported by NSF grants AST-1814440 and AST-1908570. J.T.H. is supported by NASA award 80NSSC22K0127.  W.B.H. is supported by the National Science Foundation Graduate Research Fellowship Program under Grant No.~2236415. Any opinions, findings, and conclusions or recommendations expressed in this material are those of the author(s) and do not necessarily reflect the views of the National Science Foundation.

The LBT is an international collaboration among institutions in the United States, Italy and Germany. LBT Corporation partners are: The University of Arizona on behalf of the Arizona Board of Regents; Istituto Nazionale di Astrofisica, Italy; LBT Beteiligungsgesellschaft, Germany, representing the Max-Planck Society, The Leibniz Institute for Astrophysics Potsdam, and Heidelberg University; The Ohio State University, representing OSU, University of Notre Dame, University of Minnesota and University of Virginia.  Some of the data presented herein were obtained at the W. M. Keck Observatory, which is operated as a scientific partnership among the California Institute of Technology, the University of California, and the National Aeronautics and Space Administration. The Observatory was made possible by the generous financial support of the W. M. Keck Foundation. This paper made use of the modsIDL spectral data reduction pipeline developed in part with funds provided by NSF Grant AST-1108693 and a generous gift from OSU Astronomy alumnus David G. Price through the Price Fellowship in Astronomical Instrumentation. 

This research has made use of data and/or software provided by the High Energy Astrophysics Science Archive Research Center (HEASARC), which is a service of the Astrophysics Science Division at NASA/GSFC and the High Energy Astrophysics Division of the Smithsonian Astrophysical Observatory.  This research has made use of the XRT Data Analysis Software (XRTDAS) developed under the responsibility of the ASI Science Data Center (ASDC), Italy. At Penn State the NASA \swift\ program is supported through contract NAS5-00136.  Based on observations obtained with \xmm, an ESA science mission with instruments and contributions directly funded by ESA Member States and NASA.  This research has made use of data obtained from the Chandra Data Archive and software provided by the Chandra X-ray Center (CXC) in the application packages CIAO.

Observations made with the NASA Galaxy Evolution Explorer (GALEX) were used in the analyses presented in this manuscript. Some of the data presented in this paper were obtained from the Mikulski Archive for Space Telescopes (MAST). STScI is operated by the Association of Universities for Research in Astronomy, Inc., under NASA contract NAS5-26555. Support for MAST for non-HST data is provided by the NASA Office of Space Science via grant NNX13AC07G and by other grants and contracts.

This publication makes use of data products from the Wide-field Infrared Survey Explorer, which is a joint project of the University of California, Los Angeles, and the Jet Propulsion Laboratory/California Institute of Technology, funded by the National Aeronautics and Space Administration. This publication also makes use of data products from NEOWISE, which is a project of the Jet Propulsion Laboratory/California Institute of Technology, funded by the Planetary Science Division of the National Aeronautics and Space Administration.

This work is based on observations made by ASAS-SN, UH88, and Keck. The authors wish to recognize and acknowledge the very significant cultural role and reverence that the summits of Haleakal\=a and Maunakea have always had within the indigenous Hawaiian community. We are most fortunate to have the opportunity to conduct observations from these mountains. Without their generous hospitality, the observations presented herein would not have been possible.

\section{Data availability statement}
All photometric data used in this paper are publicly available.  The optical and NIR spectra from 2022 are available upon request.  The optical spectrum from 2014 was provided by Prof. Misty C. Bentz \citep{bentz14}.

\bibliographystyle{mnras}
\bibliography{bibliography}

\begin{table*}
\caption{UV/optical photometry from \xmm, \swift, and GALEX}\label{tab:swift}
\begin{tabularx}{0.9\textwidth}{r *{1}{>{\centering\arraybackslash}X} ccccccc} \toprule
\multicolumn{1}{c}{OBSID} & MJD & $V$ & $B$ & $U$ & \textit{UVW}1 & \textit{UVM}2 & \textit{UVW}2 \\ \midrule
\multicolumn{1}{l}{\xmm} & & & & & & &   \\
0112551701 & 52439.5 & $13.65\pm0.01$ & $-$ & $-$ & $16.02\pm0.01$ & $-$ & $-$   \\
0805080401 & 57906.9 & $-$ & $-$ & $-$ & $16.39\pm0.01$ & $-$ & $-$  \\
0805080501 & 57909.3 & $-$ & $-$ & $-$ & $16.45\pm0.01$ & $-$ & $-$  \\
\multicolumn{1}{l}{\swift} & & & & & & &  \\
00080685001 & 56852.4 & $13.83\pm0.03$ & $14.39\pm0.04$ & $14.97\pm0.03$ & $15.66\pm0.04$ & $16.03\pm0.04$ & $16.14\pm0.02$   \\
00080685002 & 59361.0 & $13.97\pm0.02$ & $14.61\pm0.03$ & $15.69\pm0.04$ & $16.82\pm0.05$ & $17.34\pm0.07$ & $17.41\pm0.04$  \\
00080685003 & 59367.4 & $13.94\pm0.03$ & $14.55\pm0.03$ & $15.54\pm0.04$ & $16.48\pm0.04$ & $17.08\pm0.06$ & $17.20\pm0.04$   \\
00080685004 & 59373.4 & $13.97\pm0.03$ & $-$ & $-$ & $-$ & $-$ & $17.41\pm0.06$ \\
00080685006 & 59379.5 & $13.95\pm0.03$ & $14.62\pm0.04$ & $15.74\pm0.04$ & $16.79\pm0.06$ & $17.59\pm0.10$ & $17.54\pm0.05$  \\
00015085001 & 59663.3 & $13.86\pm0.03$ & $14.48\pm0.04$ & $15.27\pm0.04$ & $16.03\pm0.04$ & $16.55\pm0.05$ & $16.67\pm0.04$   \\
00015085002 & 59666.3 & $13.81\pm0.03$ & $14.42\pm0.04$ & $15.06\pm0.04$ & $15.75\pm0.04$ & $16.16\pm0.04$ & $16.32\pm0.04$  \\
00015085003 & 59669.8 & $13.92\pm0.03$ & $14.49\pm0.04$ & $15.33\pm0.04$ & $16.11\pm0.04$ & $16.60\pm0.04$ & $16.85\pm0.04$  \\
00015085004 & 59672.5 & $13.87\pm0.04$ & $14.54\pm0.04$ & $15.39\pm0.04$ & $16.18\pm0.05$ & $16.78\pm0.05$ & $16.87\pm0.05$  \\
00015085005 & 59675.5 & $13.85\pm0.04$ & $14.40\pm0.04$ & $15.09\pm0.04$ & $15.80\pm0.04$ & $16.25\pm0.04$ & $16.43\pm0.04$  \\
00015085006 & 59678.8 & $13.86\pm0.03$ & $14.49\pm0.04$ & $15.31\pm0.04$ & $16.06\pm0.04$ & $16.43\pm0.04$ & $16.67\pm0.04$  \\
00015085007 & 59681.0 & $13.88\pm0.03$ & $14.44\pm0.04$ & $15.16\pm0.04$ & $15.85\pm0.04$ & $16.33\pm0.04$ & $16.49\pm0.04$   \\
00015085008 & 59684.5 & $13.85\pm0.03$ & $14.33\pm0.04$ & $14.93\pm0.04$ & $15.60\pm0.04$ & $15.96\pm0.04$ & $16.18\pm0.04$  \\
00015085009 & 59688.0 & $13.72\pm0.03$ & $14.31\pm0.04$ & $14.93\pm0.04$ & $15.54\pm0.04$ & $15.94\pm0.04$ & $16.16\pm0.04$  \\
00015085010 & 59694.6 & $13.80\pm0.03$ & $14.32\pm0.03$ & $14.95\pm0.04$ & $15.60\pm0.04$ & $15.94\pm0.04$ & $16.17\pm0.04$  \\
00015085011 & 59704.3 & $13.83\pm0.03$ & $14.37\pm0.03$ & $15.01\pm0.04$ & $15.70\pm0.04$ & $16.10\pm0.04$ & $16.31\pm0.04$   \\
00015085012 & 59714.3 & $13.84\pm0.03$ & $14.40\pm0.03$ & $15.15\pm0.04$ & $15.87\pm0.04$ & $16.29\pm0.04$ & $16.51\pm0.04$   \\
00015085013 & 59724.1 & $13.80\pm0.03$ & $14.35\pm0.03$ & $15.02\pm0.04$ & $15.73\pm0.04$ & $16.19\pm0.04$ & $16.36\pm0.04$   \\
00015085014 & 59728.7 & $13.76\pm0.04$ & $14.29\pm0.04$ & $14.93\pm0.04$ & $15.61\pm0.04$ & $15.95\pm0.04$ & $16.14\pm0.04$ \\
00015085015 & 59734.3 & $13.83\pm0.03$ & $14.39\pm0.03$ & $15.03\pm0.04$ & $15.77\pm0.04$ & $16.21\pm0.04$ & $16.39\pm0.04$   \\
00015085016 & 59739.2 & $13.76\pm0.03$ & $14.27\pm0.04$ & $14.87\pm0.04$ & $15.54\pm0.04$ & $15.91\pm0.04$ & $16.09\pm0.04$   \\
00015085017 & 59744.0 & $13.77\pm0.03$ & $14.29\pm0.03$ & $14.88\pm0.04$ & $15.51\pm0.04$ & $15.92\pm0.04$ & $16.09\pm0.04$   \\
00015085019 & 59759.1 & $13.74\pm0.03$ & $14.22\pm0.04$ & $14.73\pm0.04$ & $15.41\pm0.04$ & $15.78\pm0.04$ & $15.96\pm0.04$   \\
00015085020 & 59764.7 & $13.76\pm0.03$ & $14.27\pm0.03$ & $14.86\pm0.04$ & $15.53\pm0.04$ & $15.92\pm0.04$ & $16.07\pm0.04$   \\
00015085021 & 59769.4 & $13.72\pm0.04$ & $14.27\pm0.04$ & $14.77\pm0.04$ & $15.37\pm0.04$ & $15.78\pm0.04$ & $15.93\pm0.04$  \\
00015085022 & 59774.7 & $13.82\pm0.03$ & $14.33\pm0.03$ & $14.93\pm0.04$ & $15.66\pm0.04$ & $16.05\pm0.04$ & $16.27\pm0.04$  \\
00015085023 & 59775.0 & $13.79\pm0.03$ & $14.32\pm0.04$ & $14.93\pm0.04$ & $15.61\pm0.04$ & $16.05\pm0.04$ & $16.25\pm0.04$ \\
00015085024 & 59778.4 & $-$ & $14.27\pm0.04$ & $14.85\pm0.04$ & $15.58\pm0.04$ & $-$ & $16.12\pm0.04$  \\
00015085025 & 59781.8 & $13.80\pm0.04$ & $14.30\pm0.04$ & $14.91\pm0.04$ & $15.65\pm0.04$ & $16.14\pm0.04$ & $16.38\pm0.04$ \\
00015085026 & 59784.5 & $13.86\pm0.03$ & $14.41\pm0.03$ & $15.03\pm0.04$ & $15.75\pm0.04$ & $16.23\pm0.04$ & $16.39\pm0.04$  \\
00015085027 & 59787.1 & $13.82\pm0.03$ & $14.39\pm0.03$ & $15.12\pm0.04$ & $15.90\pm0.04$ & $16.44\pm0.04$ & $16.61\pm0.04$  \\
\multicolumn{1}{l}{GALEX} & & & & & & \textit{NUV} & \textit{FUV} \\
$-$ & 53439.9 & $-$ & $-$ & $-$ & $-$ & $18.08\pm0.04$ &	$18.60\pm0.09$ \\
$-$ & 54575.6 & $-$ & $-$ & $-$ & $-$ & $18.31\pm0.01$ &	$19.04\pm0.03$ \\
$-$ & 55631.8 & $-$ & $-$ & $-$ & $-$ & $18.21\pm0.01$ & $-$  \\ \bottomrule
\end{tabularx}
\end{table*}

\begin{table*}
\caption{Optical and NIR Spectroscopy}\label{tab:spec}
\begin{tabular}{l c c c c} \toprule
\multirow{2}{*}{$\hphantom{,}$MJD} & \multirow{2}{*}{Telescope} & \multirow{2}{*}{Instrument} & Slit & Resolution$^{d}$ \\
 & & & [$''$] & [\AA] \\ \midrule
45744$^{a}$ & \multicolumn{2}{c}{Palomar 200''} & 2 & 3 \\
53858$^{b}$ & \multicolumn{2}{c}{SDSS} & 3 & 3 \\
56822$^{c}$ & \multicolumn{2}{c}{APO 3.5-m} & 5 & 15 \\ \midrule
59668 & Keck & LRIS & 1 & 7 \\
59699 &  &  &  &  \\
59813 &  &  &  &  \\
59661 & LBT & MODS & 1 & 2.5 \\
59700 &  &  &  &  \\
59649 & UH88 & SNIFS & N/A$^e$ & 6 \\
59651 &  &  &  &  \\
59700 &  &  &  &  \\
59726 &  &  &  &  \\
59769 &  &  &  &  \\
59810 &  &  &  &  \\ \midrule
55359 & IRTF & SpeX & 0.8 & 20 \\ 
59682 &  &  &  &  \\
59726 &  &  &  &  \\
59787 &  &  &  &  \\ \bottomrule
\end{tabular}
\begin{flushleft}
\textit{Notes:}  $^{a,b,c}$See \citet{ho95,ahumada20,bentz14}, respectively.  $^{d}$Resolution is approximate. $^{e}$SNIFS is an IFU. 
\end{flushleft}
\end{table*}

\begin{table*}
\caption{Broad Balmer line fluxes}\label{tab:balmer}
\begin{tabularx}{0.7\linewidth}{p{0.05\textwidth}>{\centering}p{0.08\textwidth}>{\centering}p{0.15\textwidth}>{\centering\arraybackslash}p{0.15\textwidth}>{\centering\arraybackslash}p{0.1\textwidth}} \toprule
\multirow{2}{*}{MJD} & \multirow{2}{*}{Telescope} & H$\alpha$ &  H$\beta$ & \multirow{2}{*}{H$\alpha$/H$\beta$} \\
 &  & \multicolumn{2}{c}{[$10^{-13}$ erg s$^{-1}$ cm$^{-2}$]} &   \\ \midrule
45744 & Palomar & $4.35\pm0.10$ & $0.84\pm0.25$ & $5.17\pm0.30$ \\
53858 & SDSS & $2.06\pm0.22$ & $0.34\pm0.11$ & $6.03\pm0.33$ \\
56822 & Bentz & $4.24\pm0.20$ & $1.50\pm0.23$ & $2.83\pm0.16$ \\ \midrule
59668 & Keck & $7.10\pm0.46$ & $2.75\pm0.22$ & $2.58\pm0.10$ \\
59699 &  & $8.57\pm0.32$ & $3.18\pm0.33$ & $2.70\pm0.11$ \\
59813 &  & $9.05\pm0.58$ & $2.99\pm0.17$ & $3.03\pm0.09$ \\
59661 & MODS & $5.77\pm0.65$ & $2.07\pm0.08$ & $2.78\pm0.12$ \\
59700 &  & $6.93\pm0.34$ & $2.50\pm0.03$ & $2.77\pm0.05$ \\
59649 & SNIFS & $5.56\pm0.19$ & $1.58\pm0.35$ & $3.52\pm0.23$ \\
59651 &  & $5.97\pm0.18$ & $1.88\pm0.38$ & $3.18\pm0.21$ \\
59700 &  & $5.86\pm0.30$ & $1.79\pm0.46$ & $3.27\pm0.26$ \\
59726 &  & $6.99\pm0.29$ & $2.52\pm0.27$ & $2.78\pm0.12$ \\
59769 &  & $6.60\pm1.19$ & $2.15\pm0.34$ & $3.08\pm0.24$ \\
59810 &  & $6.63\pm0.84$ & $2.27\pm0.20$ & $2.92\pm0.15$ \\ \bottomrule
\end{tabularx}
\end{table*}

\begin{table*}
\caption{X-ray fluxes and model parameters for blackbody + absorbed power-law model}\label{tab:xflux}
\begin{tabularx}{0.9\textwidth}{c c c  c c c c c } \toprule
\multicolumn{1}{c}{\multirow{2}{*}{OBSID}} & \multirow{2}{*}{MJD} &  log Soft & log Hard & log Total & $N_{\rm H}$ & \multirow{2}{*}{\textit{HR}} & \multirow{2}{*}{$\alpha_{\rm ox}$} \\ 
 &  &   & \multicolumn{1}{c}{[erg s$^{-1}$ cm$^{-2}$]} &  & [$10^{22}$ cm$^{-2}$] &  \\ \midrule
\multicolumn{1}{l}{\xmm} & & & & & &  \\
0112551701 & 52439.5 & $-11.81^{+0.01}_{-0.01}$ & $-11.19^{+0.01}_{-0.01}$ & $-11.10^{+0.01}_{-0.01}$ & $0.58^{+0.02}_{-0.02}$ & $-0.18\pm0.01$ & $1.30\pm0.01$ \\
0805080401 & 57906.9 & $-12.62^{+0.01}_{-0.01}$ & $-11.49^{+0.01}_{-0.01}$ & $-11.46^{+0.01}_{-0.01}$ & $1.61^{+0.03}_{-0.03}$ & $\hphantom{-}0.46\pm0.01$ & $1.35\pm0.01$  \\
0805080501 & 57909.3 & $-12.85^{+0.01}_{-0.01}$ & $-11.82^{+0.01}_{-0.01}$ & $-11.78^{+0.01}_{-0.01}$ & $1.78^{+0.07}_{-0.07}$ & $\hphantom{-}0.26\pm0.02$ & $1.47\pm0.01$ \\
\multicolumn{1}{l}{\swift} & & & & & & &   \\
00080685001 & 56852.4  & $-12.01^{+0.03}_{-0.03}$ & $-10.81^{+0.02}_{-0.02}$ & $-10.78^{+0.02}_{-0.02}$ & $1.84^{+0.12}_{-0.11}$ & $\hphantom{-}0.53\pm0.03$ & $1.23\pm0.03$ \\
00080685002 & 59361.0 & $-12.60^{+0.09}_{-0.07}$ & $-11.07^{+0.04}_{-0.04}$ & $-11.06^{+0.04}_{-0.04}$ & $2.79^{+0.47}_{-0.38}$ & $\hphantom{-}0.77\pm0.06$ & $1.15\pm0.05$ \\
00080685003 & 59367.4 & $-12.49^{+0.08}_{-0.07}$ & $-11.20^{+0.05}_{-0.05}$ & $-11.18^{+0.04}_{-0.05}$ & $1.85^{+0.37}_{-0.30}$ & $\hphantom{-}0.66\pm0.08$ & $1.26\pm0.06$ \\
00080685006 & 59379.5 & $-12.56^{+0.21}_{-0.10}$ & $-11.07^{+0.07}_{-0.08}$ & $-11.05^{+0.07}_{-0.08}$ & $3.00^{+0.99}_{-0.82}$ & $\hphantom{-}0.70\pm0.17$ & $1.15\pm0.09$ \\
00015085001 & 59663.3 & $-11.84^{+0.06}_{-0.05}$ & $-10.89^{+0.05}_{-0.04}$ & $-10.85^{+0.04}_{-0.04}$ & $1.27^{+0.27}_{-0.22}$ & $\hphantom{-}0.22\pm0.10$ & $1.21\pm0.05$  \\
00015085002 & 59666.3 & $-11.88^{+0.06}_{-0.05}$ & $-10.90^{+0.04}_{-0.04}$ & $-10.86^{+0.04}_{-0.04}$ & $1.51^{+0.27}_{-0.23}$ & $\hphantom{-}0.24\pm0.09$ & $1.26\pm0.05$ \\
00015085003 & 59669.8  & $-12.49^{+0.08}_{-0.07}$ & $-11.23^{+0.05}_{-0.06}$ & $-11.21^{+0.05}_{-0.05}$ & $1.79^{+0.35}_{-0.29}$ & $\hphantom{-}0.63\pm0.08$ & $1.33\pm0.06$ \\
00015085004 & 59672.5  & $-12.49^{+0.17}_{-0.11}$ & $-11.33^{+0.08}_{-0.09}$ & $-11.30^{+0.08}_{-0.09}$ & $2.29^{+1.78}_{-0.81}$ & $\hphantom{-}0.38\pm0.20$ & $1.35\pm0.15$ \\
00015085005 & 59675.5 & $-11.82^{+0.06}_{-0.05}$ & $-10.82^{+0.04}_{-0.04}$ & $-10.77^{+0.04}_{-0.03}$ & $1.15^{+0.22}_{-0.18}$ & $\hphantom{-}0.37\pm0.09$ & $1.22\pm0.05$ \\
00015085006 & 59678.8  & $-11.84^{+0.04}_{-0.04}$ & $-10.92^{+0.04}_{-0.03}$ & $-10.87^{+0.03}_{-0.04}$ & $1.10^{+0.21}_{-0.18}$ & $\hphantom{-}0.21\pm0.08$ & $1.22\pm0.05$ \\
00015085007 & 59681.0 & $-11.89^{+0.05}_{-0.05}$ & $-10.94^{+0.05}_{-0.05}$ & $-10.89^{+0.04}_{-0.04}$ & $1.15^{+0.28}_{-0.23}$ & $\hphantom{-}0.27\pm0.09$ & $1.26\pm0.06$ \\
00015085008 & 59684.5 & $-11.99^{+0.06}_{-0.05}$ & $-10.85^{+0.04}_{-0.03}$ & $-10.82^{+0.03}_{-0.04}$ & $2.03^{+0.31}_{-0.27}$ & $\hphantom{-}0.39\pm0.08$ & $1.26\pm0.05$  \\
00015085009 & 59688.0 & $-11.87^{+0.05}_{-0.04}$ & $-10.84^{+0.03}_{-0.03}$ & $-10.80^{+0.03}_{-0.03}$ & $1.48^{+0.22}_{-0.19}$ & $\hphantom{-}0.33\pm0.07$ & $1.27\pm0.04$ \\
00015085010 & 59694.6 & $-11.40^{+0.02}_{-0.02}$ & $-10.73^{+0.02}_{-0.02}$ & $-10.65^{+0.02}_{-0.02}$ & $0.84^{+0.09}_{-0.08}$ & $-0.17\pm0.03$ & $1.22\pm0.03$ \\
00015085011 & 59704.3 & $-11.63^{+0.03}_{-0.02}$ & $-10.94^{+0.03}_{-0.03}$ & $-10.86^{+0.02}_{-0.02}$ & $0.75^{+0.11}_{-0.10}$ & $-0.11\pm0.05$ & $1.29\pm0.03$ \\
00015085012 & 59694.6  & $-12.28^{+0.05}_{-0.04}$ & $-11.02^{+0.03}_{-0.03}$ & $-11.00^{+0.03}_{-0.03}$ & $2.45^{+0.29}_{-0.25}$ & $\hphantom{-}0.50\pm0.06$ & $1.28\pm0.04$ \\
00015085013 & 59724.1 & $-12.07^{+0.04}_{-0.04}$ & $-11.00^{+0.03}_{-0.03}$ & $-10.97^{+0.02}_{-0.03}$ & $1.42^{+0.18}_{-0.16}$ & $\hphantom{-}0.41\pm0.05$ & $1.30\pm0.04$ \\
00015085014 & 59728.7 & $-11.95^{+0.07}_{-0.06}$ & $-10.75^{+0.04}_{-0.04}$ & $-10.73^{+0.04}_{-0.04}$ & $1.95^{+0.32}_{-0.27}$ & $\hphantom{-}0.49\pm0.08$ & $1.22\pm0.05$ \\
00015085015 & 59734.3 & $-12.21^{+0.05}_{-0.05}$ & $-11.06^{+0.04}_{-0.03}$ & $-11.03^{+0.03}_{-0.03}$ & $1.78^{+0.30}_{-0.26}$ & $\hphantom{-}0.46\pm0.07$ & $1.32\pm0.05$ \\
00015085016 & 59739.2 & $-11.89^{+0.04}_{-0.04}$ & $-10.82^{+0.03}_{-0.03}$ & $-10.78^{+0.03}_{-0.03}$ & $1.84^{+0.24}_{-0.21}$ & $\hphantom{-}0.32\pm0.07$ & $1.26\pm0.04$ \\
00015085017 & 59744.0 & $-11.85^{+0.03}_{-0.03}$ & $-10.77^{+0.02}_{-0.02}$ & $-10.74^{+0.02}_{-0.02}$ & $1.64^{+0.16}_{-0.15}$ & $\hphantom{-}0.37\pm0.05$ & $1.25\pm0.03$ \\
00015085019 & 59759.1 & $-12.15^{+0.06}_{-0.06}$ & $-10.93^{+0.04}_{-0.04}$ & $-10.90^{+0.04}_{-0.04}$ & $2.03^{+0.35}_{-0.30}$ & $\hphantom{-}0.52\pm0.07$ & $1.32\pm0.05$  \\
00015085020 & 59764.7 & $-11.90^{+0.04}_{-0.03}$ & $-10.79^{+0.02}_{-0.03}$ & $-10.76^{+0.02}_{-0.02}$ & $1.40^{+0.14}_{-0.13}$ & $\hphantom{-}0.48\pm0.05$ & $1.25\pm0.03$  \\
00015085021 & 59769.4 & $-11.47^{+0.04}_{-0.04}$ & $-10.69^{+0.04}_{-0.04}$ & $-10.62^{+0.04}_{-0.04}$ & $1.02^{+0.20}_{-0.17}$ & $-0.02\pm0.08$ & $1.24\pm0.05$ \\
00015085022 & 59774.7 & $-11.82^{+0.04}_{-0.03}$ & $-10.91^{+0.03}_{-0.03}$ & $-10.86^{+0.03}_{-0.03}$ & $1.08^{+0.17}_{-0.15}$ & $\hphantom{-}0.21\pm0.06$ & $1.28\pm0.04$ \\
00015085023 & 59775.0 & $-11.67^{+0.04}_{-0.04}$ & $-10.76^{+0.04}_{-0.04}$ & $-10.71^{+0.04}_{-0.04}$ & $1.25^{+0.21}_{-0.18}$ & $\hphantom{-}0.15\pm0.08$ & $1.23\pm0.05$ \\
00015085024 & 59778.4  & $-11.67^{+0.05}_{-0.04}$ & $-10.65^{+0.04}_{-0.04}$ & $-10.61^{+0.03}_{-0.04}$ & $1.83^{+0.24}_{-0.20}$ & $\hphantom{-}0.24\pm0.09$ & $1.19\pm0.04$ \\
00015085025 & 59781.8  & $-11.72^{+0.05}_{-0.04}$ & $-10.75^{+0.04}_{-0.04}$ & $-10.71^{+0.03}_{-0.04}$ & $1.56^{+0.27}_{-0.23}$ & $\hphantom{-}0.20\pm0.08$ & $1.22\pm0.05$ \\
00015085026 & 59784.5 & $-12.08^{+0.05}_{-0.04}$ & $-10.88^{+0.03}_{-0.03}$ & $-10.85^{+0.03}_{-0.03}$ & $2.15^{+0.30}_{-0.26}$ & $\hphantom{-}0.46\pm0.06$ & $1.24\pm0.04$ \\
00015085027 & 59787.1  & $-12.40^{+0.07}_{-0.07}$ & $-11.14^{+0.04}_{-0.04}$ & $-11.12^{+0.04}_{-0.04}$ & $2.50^{+0.43}_{-0.35}$ & $\hphantom{-}0.50\pm0.08$ & $1.32\pm0.05$ \\
\multicolumn{1}{l}{\nicer} & & & & & &  \\
5202700101 & 59670.3 & $-12.37^{+0.02}_{-0.02}$ & $-11.22^{+0.02}_{-0.02}$ & $-11.19^{+0.02}_{-0.02}$ & $1.95^{+0.14}_{-0.13}$ & $\hphantom{-}0.42\pm0.03$ & $-$ \\
5202700102 & 59688.8  & $-11.56^{+0.01}_{-0.01}$ & $-10.63^{+0.01}_{-0.01}$ & $-10.58^{+0.01}_{-0.01}$ & $1.41^{+0.07}_{-0.07}$ & $\hphantom{-}0.16\pm0.02$ & $-$ \\
5202700103 & 59700.8 & $-11.12^{+0.01}_{-0.01}$ & $-10.55^{+0.02}_{-0.02}$ & $-10.45^{+0.01}_{-0.01}$ & $0.78^{+0.05}_{-0.05}$ & $-0.30\pm0.02$ & $-$ \\
5202700104 & 59715.1 & $-12.13^{+0.02}_{-0.02}$ & $-10.90^{+0.02}_{-0.02}$ & $-10.88^{+0.02}_{-0.02}$ & $2.81^{+0.20}_{-0.19}$ & $\hphantom{-}0.40\pm0.02$ & $-$ \\
5202700105 & 59763.8 & $-12.04^{+0.01}_{-0.01}$ & $-10.89^{+0.01}_{-0.01}$ & $-10.86^{+0.01}_{-0.01}$ & $1.92^{+0.05}_{-0.05}$ & $\hphantom{-}0.42\pm0.01$ & $-$ \\
5202700106 & 59764.0 & $-11.99^{+0.01}_{-0.01}$ & $-10.85^{+0.01}_{-0.01}$ & $-10.82^{+0.01}_{-0.01}$ & $1.86^{+0.06}_{-0.06}$ & $\hphantom{-}0.43\pm0.01$ & $-$ \\
\multicolumn{1}{l}{\chandra} & & & & & & &  \\
415 & 51790.7 & $-13.12^{+0.08}_{-0.07}$ & $-11.86^{+0.06}_{-0.06}$ & $-11.83^{+0.06}_{-0.06}$ & $3.64^{+1.07}_{-0.79}$ & $\hphantom{-}0.38\pm0.09$ & $-$ \\
\multicolumn{1}{l}{\suzaku} & & & & & &  \\
708001010 & 56489.2 & $-12.18^{+0.01}_{-0.01}$ & $-11.07^{+0.01}_{-0.01}$ & $-11.04^{+0.01}_{-0.01}$ & $1.98^{+0.03}_{-0.03}$ & $\hphantom{-}0.36\pm0.01$ & $-$ \\ \bottomrule
\end{tabularx}
\begin{flushleft}
\textit{Notes:}  See Sec.~\ref{sec:xflux} for definitions of column headers.
\end{flushleft}
\end{table*}

\begin{table*}
\caption{X-ray fluxes and model parameters for ionized absorption model}\label{tab:xparams}
\begin{tabularx}{0.8\textwidth}{r *{1}{>{\centering\arraybackslash}X} c c c c c c} \toprule
\multicolumn{1}{c}{\multirow{2}{*}{OBSID}} & \multirow{2}{*}{MJD}  & \multirow{2}{*}{log $\xi$} & $N_{\rm H}$ & \multirow{2}{*}{$\Gamma$} & log Soft & log Hard \\ 
 &  &  & [$10^{22}$ cm$^{-2}$] & & \multicolumn{2}{c}{[erg s$^{-1}$ cm$^{-2}$]} \\ \midrule
\multicolumn{1}{l}{\nicer}    \\
5202700101 & 59670.3 & $\hphantom{-}0.40^{+0.09}_{-0.07}$ & $5.14^{+0.34}_{-0.36}$ & $2.07^{+0.08}_{-0.08}$ & $-12.35^{+0.01}_{-0.02}$ & $-11.23^{+0.02}_{-0.02}$ \\
5202700102 & 59688.8 & $\hphantom{-}1.36^{+0.06}_{-0.05}$ & $4.32^{+0.32}_{-0.35}$ & $1.59^{+0.05}_{-0.06}$ & $-11.56^{+0.01}_{-0.01}$ & $-10.61^{+0.01}_{-0.02}$ \\
5202700103 & 59700.8 & $\hphantom{-}1.46^{+0.07}_{-0.05}$ & $3.19^{+0.41}_{-0.27}$ & $2.06^{+0.05}_{-0.06}$ & $-11.10^{+0.01}_{-0.01}$ & $-10.65^{+0.03}_{-0.03}$ \\
5202700104 & 59715.1 & $\hphantom{-}1.51^{+0.22}_{-0.14}$ & $7.18^{+0.24}_{-0.27}$ & $1.38^{+0.12}_{-0.14}$ & $-12.10^{+0.01}_{-0.08}$ & $-10.82^{+0.01}_{-0.06}$ \\
5202700105 & 59763.8 & $\hphantom{-}1.12^{+0.01}_{-0.01}$ & $4.52^{+0.23}_{-0.27}$ & $1.63^{+0.05}_{-0.06}$ & $-12.03^{+0.01}_{-0.01}$ & $-10.88^{+0.01}_{-0.02}$ \\
5202700106 & 59764.0 & $\hphantom{-}1.19^{+0.03}_{-0.02}$ & $4.43^{+0.30}_{-0.30}$ & $1.46^{+0.06}_{-0.06}$ & $-11.99^{+0.01}_{-0.01}$ & $-10.79^{+0.01}_{-0.02}$ \\
\multicolumn{1}{l}{\xmm} \\
0112551701 & 52439.5 & $\hphantom{-}1.47^{+0.05}_{-0.05}$ & $2.09^{+0.12}_{-0.13}$ & $1.70^{+0.03}_{-0.03}$ & $-11.81^{+0.01}_{-0.01}$ & $-11.21^{+0.01}_{-0.01}$ \\
0805080401 & 57906.9 & $\hphantom{-}0.35^{+0.02}_{-0.02}$ & $3.90^{+0.13}_{-0.14}$ & $2.00^{+0.04}_{-0.04}$ & $-12.64^{+0.01}_{-0.01}$ & $-11.50^{+0.01}_{-0.01}$ \\
0805080501 & 57909.3 & $-0.54^{+0.07}_{-0.06}$ & $4.85^{+0.17}_{-0.21}$ & $2.29^{+0.04}_{-0.04}$ & $-12.85^{+0.01}_{-0.02}$ & $-11.81^{+0.01}_{-0.01}$ \\
 \bottomrule
\end{tabularx}
\begin{flushleft}
\textit{Notes:}  See Sec.~\ref{sec:xipcf} for definitions of column headers.
\end{flushleft}
\end{table*}


\label{lastpage}
\bsp	
\end{document}